\documentclass[a4paper,aps,prx,reprint,superscriptaddress]{revtex4-1}

\usepackage{amsmath} 
\usepackage[]{changes}
\usepackage{graphicx}
\usepackage{changes}
\usepackage[colorlinks,
	linkcolor=red,
	citecolor=blue,
	urlcolor=red]{hyperref}

\usepackage{lipsum}

\usepackage{mathtools} 
\newcommand{\eqdef}{\vcentcolon=\,}

\newcommand{\fref}[1]{Fig.~\ref{#1}}
\renewcommand{\eqref}[1]{Eq.~(\ref{#1})}
\newcommand{\secref}[1]{Sec.~\ref{#1}}

\newcommand{\tr}{\mathrm{Tr}}
\newcommand{\avg}[1]{\left\langle{#1}\right\rangle}

\newcommand{\re}{\mathrm{Re}\,}

\newcommand{\abs}[1]{\left\vert{#1}\right\vert}

\renewcommand{\t}[1]{\mathrm{#1}}

\begin{document}

\title{Appearance and disappearance of quantum correlations\\ 
	in measurement-based feedback control of a mechanical oscillator}

\author{V. Sudhir} 
\author{D.J. Wilson}
\author{R. Schilling} 
\author{H. Sch\"{u}tz}
\author{S.A.~Fedorov}
\author{A.H. Ghadimi}
\affiliation{Institute for Condensed Matter Physics, {\'E}cole Polytechnique F{\'e}d{\'e}rale de Lausanne, 
	Lausanne 1015, Switzerland}
\author{A. Nunnenkamp}
\affiliation{Cavendish Laboratory, University of Cambridge, Cambridge CB3 0HE, United Kingdom}
\author{T.J. Kippenberg}
\affiliation{Institute for Condensed Matter Physics, {\'E}cole Polytechnique F{\'e}d{\'e}rale de Lausanne, 
	Lausanne 1015, Switzerland}

\begin{abstract}
Quantum correlations between imprecision and back-action are a hallmark of continuous linear measurements. 
Here we study how measurement-based feedback can be used to improve the visibility of 
quantum correlations due to
the interaction of a laser field with a nano-optomechanical system.
Back-action imparted by the meter laser, due to radiation pressure quantum fluctuations, 
gives rise to correlations between its phase and amplitude quadratures. 
These quantum correlations are observed in the experiment both as squeezing of the meter field fluctuations below 
the vacuum level in a homodyne measurement, and as sideband asymmetry in a heterodyne measurement, 
demonstrating the common origin of both phenomena. 
We show that quantum feedback, i.e. feedback that suppresses measurement back-action, 
can be used to increase the visibility of the sideband asymmetry ratio.
In contrast, by operating the feedback loop in the regime of noise “squashing”, 
where the in-loop photocurrent variance is reduced below the vacuum level, the visibility of 
the sideband asymmetry is reduced. This is due to feedback back-action arising from vacuum noise in the homodyne detector.
These experiments demonstrate the possibility, as well as the
fundamental limits of measurement-based feedback as a tool to manipulate quantum correlations.
\end{abstract}

\date{\today}

\maketitle


Measurements proceed by establishing correlations between a system and a meter. 
In a quantum description of this process \cite{WisMil10}, 
the effect of measurement persists in the system in the form of measurement back-action.
For continuous linear measurement \cite{BragKhal,Clerk10},
where the meter
couples linearly and weakly to the system, correlations between the system and meter additionally manifest as 
back-action-induced quantum correlations between the degrees of freedom of the meter.
A paradigmatic example is 
the interferometric position readout of a mechanical
oscillator \cite{Abbo016a}. 
The meter in this case is an optical field, which possesses two degrees of freedom (quadratures):
amplitude and phase. 
The position of the oscillator is imprinted onto the phase quadrature.  Back-action arises from 
vacuum 
fluctuations of the amplitude quadrature, which are imprinted onto the phase via the
back-action-driven motion of the oscillator, leading to amplitude-phase quantum correlations in the meter field.
In a homodyne detector, these quantum correlations manifest as ponderomotive squeezing 
of an appropriately chosen field quadrature \cite{Stamp12,Pain13,PurReg13,Niel16}. 
In a heterodyne detector, 
they manifest as motional sideband asymmetry \cite{AmirPain12,Wein14,PurReg15,UndHarr15}. 
Differences between these effects arise from the details of
how meter fluctuations are converted to a classical signal
by the detection process \cite{Shap85,KhalPain12,Wein14,Buch16}
\footnote{Note that sideband asymmetry arising for direct photon counting of the meter field
has a different origin \cite{Wein14}.}

Here we investigate the effect of measurement-based feedback on quantum correlations due to the interaction of an 
optical field with a nano-mechanical oscillator. 
Recent advances \cite{WilSudKip15} have enabled operation of an optomechanical system such that the mechanical 
oscillator can be measured at a 
rate approaching the thermal decoherence rate, a regime where measurement back-action becomes relevant compared to 
the thermal noises.
Harnessing this capability, here we show that feedback of a homodyne measurement can be used
to improve the visibility of motional sideband asymmetry by suppressing measurement back-action.  
Indeed, the feedback loop cools the oscillator to a final phonon occupancy ($n_{\mathrm{eff}}$) that is more than two orders of lower 
than that due to the quantum backaction ($n_{\mathrm{qba}}$) of the meter beam. 
The system therefore operates in the \textit{quantum feedback} regime, where quantum back-action is effectively suppressed 
by feedback, and feedback can manipulate quantum correlations without destroying them.
This is possible because the measurement used for feedback contains a faithful record of its own back-action \cite{Hat13}.
Further we study how these quantum correlations are obscured in the regime where
feedback is dominated by quantum noise in the in-loop detector (i.e. feedback back-action); a regime giving rise to ``squashing''
of the in-loop photocurrent \cite{Taub95}. 
This demonstrates the complementary scenario where
feedback is detrimental to the observation of quantum correlations.
Conceptually, this feedback back-action dominated regime is analogous to the quantum back-action limit of sideband 
cooling \cite{PetReg15}.
Finally, we probe quantum correlations via a homodyne detector tuned close to the amplitude quadrature, 
and observe squeezing, i.e. a reduction of the homodyne noise below the vacuum level. By observing both squeezing 
and sideband asymmetry in the same device, the common origin of motional sideband asymmetry \cite{KhalPain12,Wein14} and 
optical squeezing \cite{PurReg15,Niel16} in general dyne detection \cite{WisMil10}
of the meter field is experimentally illustrated.

A pedagogical description of continuous linear measurement is germane to understanding our approach.  
We denote as $x(t)$ the position of a quantum harmonic oscillator and $y(t)\propto x(t)$ the output of a linear 
continuous position detector.  Since it is a continuous observable, $y(t)$ must commute with itself at different times 
($[y(t),y(t')]=0$).  $x(t)$ does not obey this constraint, which requires that the detector output contains an 
additional noise term $x_\t{n}(t)$ that enforces the commutator.  $x_\t{n}$ contains two components: an 
apparent (imprecision) noise, $x_\t{imp}$, which arises from quantum fluctuations of the meter degree of freedom that 
is coupled to the detector, and a physical (back-action) noise, $x_\t{ba}$, which arises from quantum fluctuations of 
the meter degree of freedom that is coupled to the system.
The total detector signal, $y=x+x_\t{ba}+x_\t{imp}\equiv x_\t{tot}+x_\t{imp}$, is characterized by a 
(symmetrized, double-sided \footnote{Here, heterodyne spectra are expressed as double-sided symmetrized
	spectra, for example $\bar{S}_{yy}$, while homodyne spectra are expressed as the corresponding single-sided
	symmetrized versions, for example $\bar{S}_y$}) noise spectrum 
\cite{Clerk10,suppinfo},
\begin{equation}\label{eq:Syy}
	\bar{S}_{yy}(\Omega) = \bar{S}_{xx}^\t{imp}(\Omega) + \bar{S}_{xx}^\t{tot}(\Omega) 
		+ 2\re \bar{S}_{x_\t{ba}x_\t{imp}}(\Omega),
\end{equation} 
which contains terms due to quantum fluctuations of the meter ($x_\t{imp}$), total physical motion ($x_\t{tot}$), 
and quantum (imprecision-back-action) correlations, respectively.

\begin{figure}[t!]
	\includegraphics[width=0.95\columnwidth]{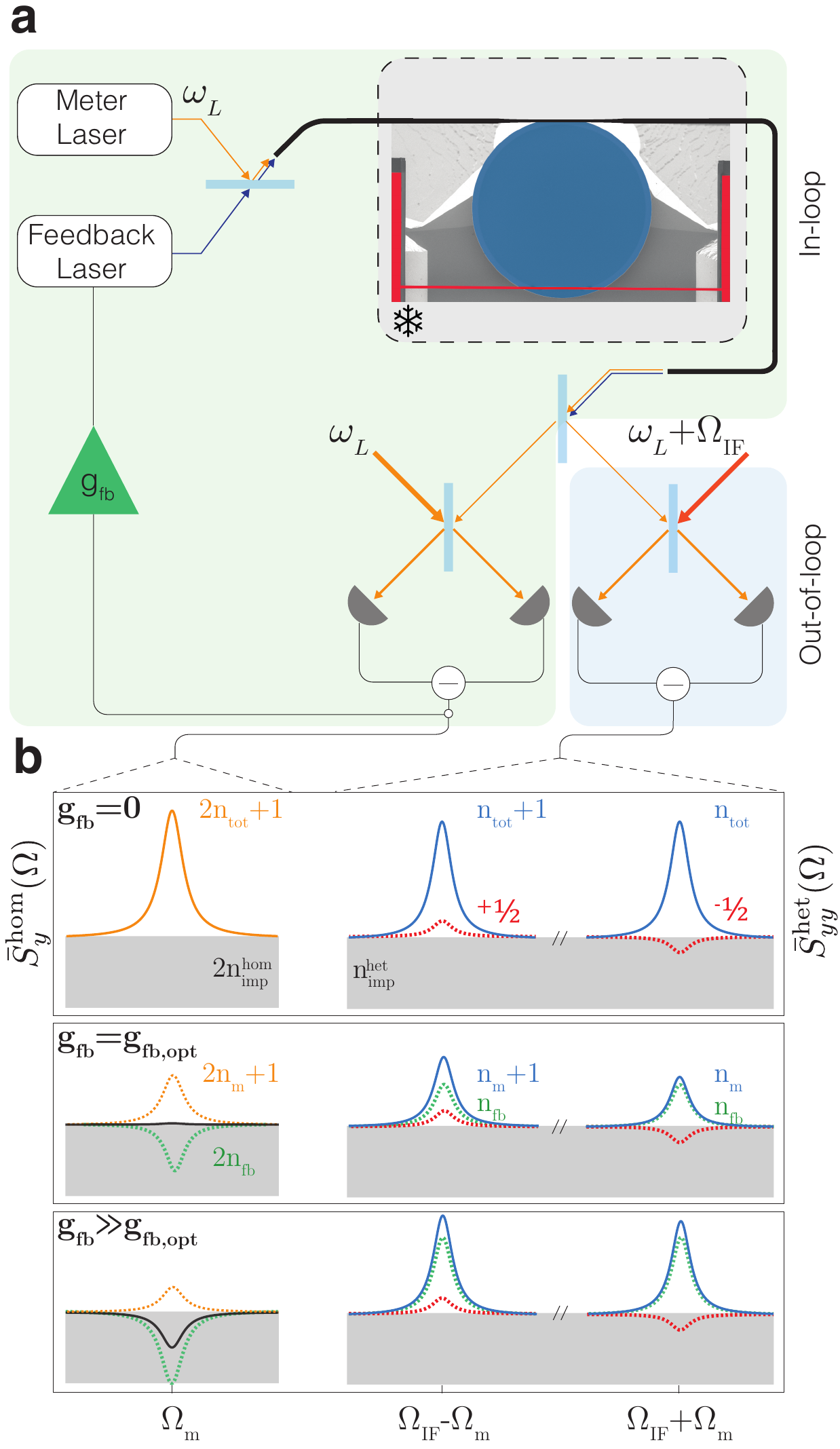}
	\caption{\label{fig1}\textbf{Using homodyne feedback to increase the visibility of 
			quantum-correlation-induced motional sideband asymmetry.} 
		(a) Linear position measurement and feedback control 
		of a nanomechanical string ($\t{Si}_\t{3}\t{N}_\t{4}$, red) is provided by evanescent coupling to 
		an optical microdisk cavity ($\t{SiO}_\t{2}$, blue). Whispering gallery modes of the microdisk are driven 
		by a pair of tunable diode lasers using a tapered optical fiber (black). The `meter' field (orange) is 
		directed to a pair of balanced interferometers (homodyne, green; heterodyne, blue).  
		A delayed and an amplified copy of the homodyne signal is imprinted onto the amplitude of the `feedback' 
		field (blue), effecting cold damping of the fundamental beam mode. Taper, nanobeam, and microdisk are 
		integrated into a He cryostat (grey). (b) Schematic of the closed-loop homodyne (left) and heterodyne 
		(right) noise spectrum for various feedback gains.  Contributions from measurement imprecision, physical 
		motion, and imprecision-back-action correlations are delineated by color.
		}
\end{figure}

In our experiment we monitor the position fluctuations of a cryogenically pre-cooled ($T \approx 6\,\t{K}$)
nanomechanical string coupled dispersively to an
optical microcavity \cite{Schil15}.
The fundamental mode of the string forms the oscillator (frequency $\Omega_\t{m} = 2\pi\cdot
4.3\;\t{MHz}$, damping rate $\Gamma_m = 2\pi \cdot 7\, \t{Hz}$).  
The meter is a laser field passing resonantly through the cavity (wavelength, $\lambda\approx 774\,\t{nm}$), 
whose quadratures
are monitored simultaneously by a homodyne and a heterodyne detector (\fref{fig1}a).  
Both detectors are operated with an imprecision far below that at the SQL, 
implying that quantum back-action due to the measurement (quantified as a phonon occupancy $n_\t{qba}$) is a significant 
contribution to the total motion of the nanomechanical oscillator ($n_\t{tot}$), i.e., $n_{\mathrm{qba}}\approx 0.15\, n_\t{tot}$.

\begin{figure}[t!]
	\includegraphics[width=0.95\columnwidth]{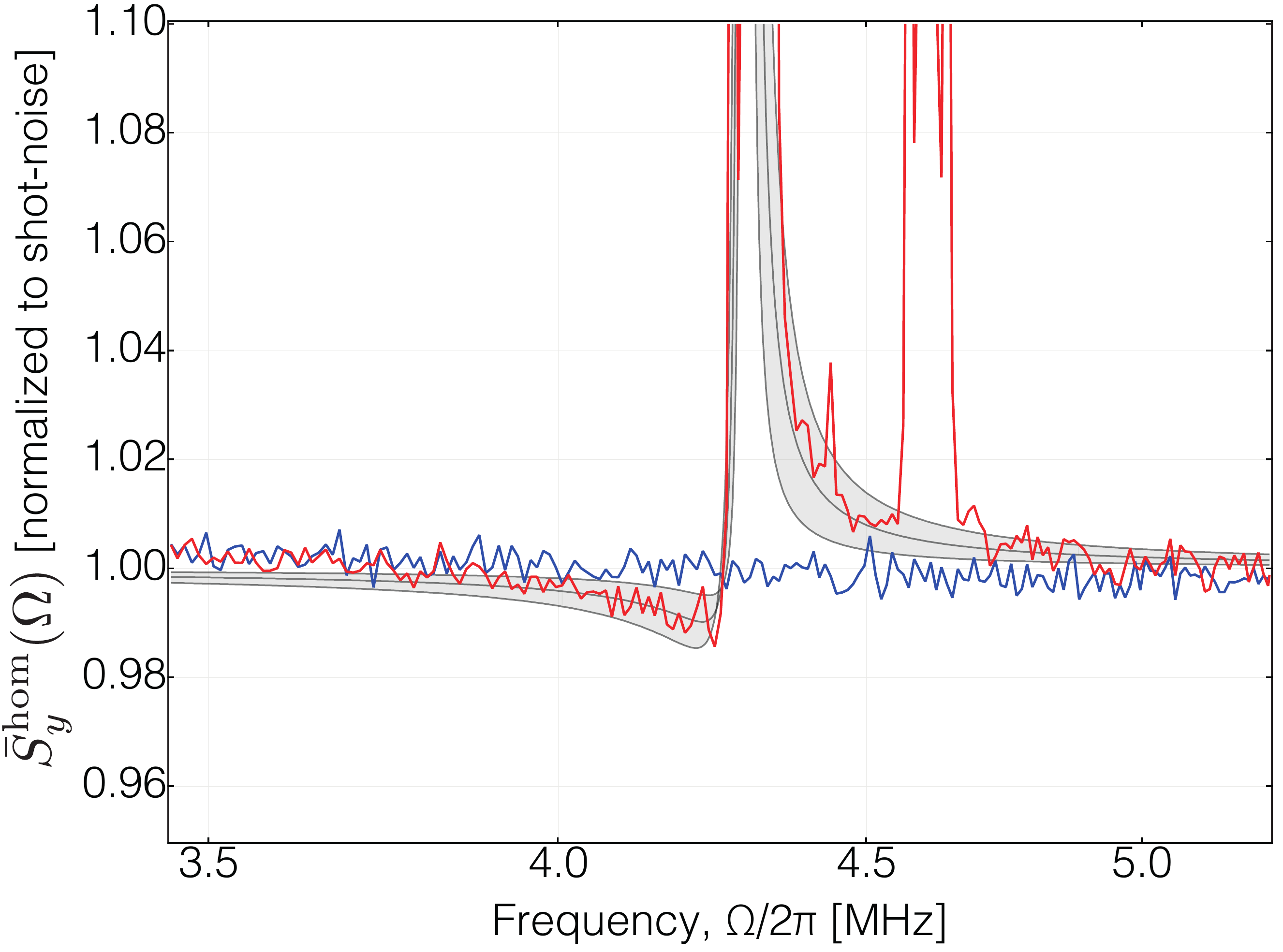}
	\caption{\label{figSq}
		\textbf{Squeezing in homodyne detection.}
		Quantum correlations in the cavity transmission manifest as photocurrent squeezing
		when measured using a homodyne detector set to near the amplitude quadrature (here $\theta \approx 0.1\, \t{rad})$. 
		Blue trace shows the shot-noise in the homodyne detector when the meter field is in vacuum. Red shows measurement
		when the meter field has interacted with the mechanical oscillator; squeezing at the level of $1\%$ is visible,
		consistent with theoretical predictions (black) using a model incorporating $\approx 5\%$ uncertainties (gray regions) 
		in the experimentally measured system parameters. The wideband shot noise detected far from mechanical resonance
		agrees very well with the expected local oscillator shot noise. By directing all of the cavity transmission 
		to the homodyne detector, we realize an overall detection efficiency, $\eta_\t{hom}\approx 0.2$.
	}	
\end{figure}

We first assess the resulting optomechanical quantum correlations by measuring the output field in a homodyne detector.
Measuring the quadrature of the meter field at phase $\theta$, imprecision-back-action
correlations in a homodyne detector take the form \cite{suppinfo},
\begin{equation}
	\bar{S}_{x_\t{ba}x_\t{imp}}^\t{hom}(\Omega)\propto C\eta_\t{hom} \sin(2\theta) \chi_m(\Omega),
\end{equation}
where $C=4g_0^2 n_c/\kappa \Gamma_m$ is the multi-photon cooperativity of the optomechanical system,
$\eta_\t{hom}$ is the detection efficiency and $\chi_m(\Omega) = (-\Omega^2 +\Omega_m^2 -i\Omega \Gamma_m)^{-1}/m$ is
the susceptibility of the mechanical oscillator to an applied force.
In the phase quadrature ($\theta=\pi/2$), where sensitivity to mechanical motion is largest (shown in \fref{fig1}b top left), 
these correlations
do not appear in the homodyne photocurrent. However, near the amplitude quadrature, $\theta\rightarrow 0$, the
magnitude of the correlation term can be comparable to the thermal motion, leading to observable squeezing
of the homodyne photocurrent \cite{Fabre94}.
\fref{figSq} shows homodyne detection of optical squeezing near the amplitude quadrature. The observed squeezing, 
while small in magnitude $1\%$, can still be clearly observed in the measurement.

\begin{figure*}[t!]
	\includegraphics[scale=0.72]{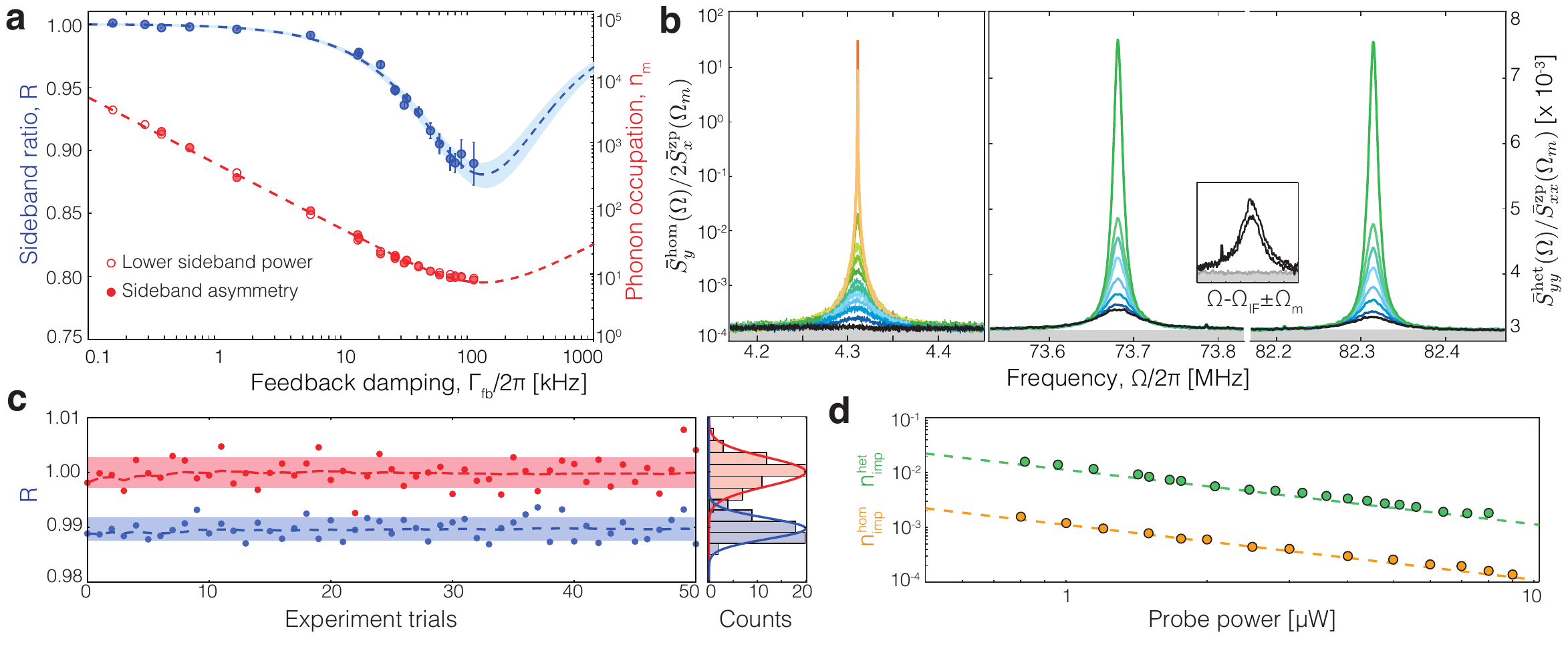}
	\caption{\label{fig2}\textbf{Motional sideband asymmetry in the heterodyne measurement of a 
	cold-damped mechanical oscillator.}
		(a) Heterodyne sideband asymmetry ($R$, blue) and inferred mechanical mode occupation ($n_\t{m}$, red) 
		versus closed-loop mechanical damping rate ($\Gamma_\t{fb}$) for various feedback gains.  
		A maximum asymmetry of $1-R\approx 12\%$ ($n_\t{m}\approx7.3$) appears as the feedback 
		gain approaches its optimal value. 
		Dashed lines correspond to models $R=\tfrac{n_\t{m}}{n_\t{m}+1}$ (\eqref{eq:R}, blue line) 
		and $n_\t{m}+\tfrac{1}{2}\approx 
		\tfrac{\Gamma_\t{m}}{\Gamma_\t{fb}}n_\t{tot}+\tfrac{\Gamma_\t{fb}}{\Gamma_\t{m}}n_\t{imp}^\t{hom}$ 
		(\eqref{eq:colddamping}, red line). Solid blue band is a confidence interval based on uncertainties in 
		estimates of $n_\t{tot}$, $n_\t{imp}^\t{hom}$, and $\Gamma_\t{m}$.
		Open red circles are independent estimates of $n_\t{m}$ based on the area beneath the left heterodyne sideband.
		(b) Homodyne (left panel) and heterodyne (right panel) spectra used to obtain (a). 
		Black traces correspond to lowest occupation; asymmetry is highlighted in the inset. 
		Only a subset of heterodyne spectra are shown, for low $n_\t{m}$, with colors matching the
		corresponding homodyne spectra. 
		An important feature of these spectra are their low imprecision, 
		$n_\t{imp}^\t{hom}= (16\eta_\t{hom}C_0 n_\t{c})^{-1} = 1.2\cdot 10^{-4}$ and 
		$n_\t{imp}^\t{het} =(4\eta_\t{het}C_0 n_c)^{-1} = 2.9\cdot 10^{-3}$.
		This is made possible by the high photon collection efficiency $\eta\sim0.2$, single photon cooperativity 
		$C_\t{0}=4g_0^2/\kappa\Gamma_\t{m}=0.3$, and power handling capacity of the microcavity-based sensor 
		(allowing for intracavity photon numbers of $n_\t{c}\sim 10^4$).
		(c) Statistical fluctuations of $R$ for low feedback gain, indicating the ability to discriminate 
		a $0.5\%$ asymmetry, corresponding to $n_\t{m}\approx 100$. 
		(d) Phonon-equivalent imprecision of the heterodyne and homodyne detectors as a function of the power 
		of the meter field. 
	}
\end{figure*}

Detecting ponderomotive squeezing provides bona fide proof of the presence of quantum correlations in the meter field. We next
probe the alternate manifestation of these correlations as sideband asymmetry -- in a heterodyne detector.
The heterodyne detector used in the experiment monitors both quadratures of the meter
simultaneously \cite{suppinfo}, giving access to $\bar{S}_{yy}^\t{het}(\Omega>0)$,
where $\bar{S}_{yy}^\t{het}(\Omega_\t{IF}\pm\Omega_\t{m})$ correspond to upper ($+$) and 
lower ($-$) motional sidebands (displaced by the heterodyne intermediate frequency, $\Omega_\t{IF}$).
Quantum correlations between the phase and amplitude of the meter manifest as an asymmetry of the heterodyne motional 
sidebands. This can be understood from the three terms in \eqref{eq:Syy}, illustrated as components of the 
heterodyne signal in \fref{fig1}b (top right panel).
Detector imprecision (gray) -- arising from the vacuum 
fluctuations in the phase and amplitude quadrature of the probe -- contributes a phonon-equivalent noise of
$n_\t{imp}^\t{het}\equiv \bar{S}_{yy}^\t{het,imp}(\Omega_\t{IF}\pm\Omega_\t{m})/\bar{S}_{xx}^\t{zp}(\Omega_m)$.
Physical motion -- arising from a combination of thermal force and meter back-action -- 
contributes $n_\t{m}+\tfrac{1}{2}$ phonons to each sideband. 
Imprecision-back-action correlations --
arising from amplitude-phase correlations in the meter --
contribute $\pm\tfrac{1}{2}$ phonons to the lower/upper 
sideband (red dashed) \cite{suppinfo}, where $\bar{S}_{xx}^\t{zp}(\Omega_m)=\frac{4x_\t{zp}^2}{\Gamma_m}$
is the zero-point position spectral density on resonance. 
The resulting asymmetry of the sidebands (blue traces),
\begin{equation}\label{eq:R}
R \equiv \frac{\bar{S}_{yy}^\t{het}(\Omega_\t{het}^{+})- 
	\bar{S}_{yy}^\t{het,imp}(\Omega_\t{het}^{+})}{\bar{S}_{yy}^\t{het}(\Omega_\t{het}^{-})- 
	\bar{S}_{yy}^\t{het,imp}(\Omega_\t{het}^{-})}\approx \frac{n_\t{m}}{n_\t{m}+1},
\end{equation}
is commensurate with one phonon 
and arises purely from quantum correlations in the meter
(here $\Omega_\t{het}^{\pm}\equiv\Omega_\t{IF}\pm\Omega_\t{m}$). 
This asymmetry corresponds directly to the visibility of 
imprecision-back-action correlations with respect to the total noise power, i.e.,
\begin{equation}\label{eq:xihet}
\xi \equiv 
	\frac{2\re\,\bar{S}_{x_\t{ba}x_\t{imp}}(\Omega_\t{het}^+)}
	{\bar{S}_{xx}^\t{imp}(\Omega_\t{het}^+)+S_{xx}^\t{tot}(\Omega_\t{het}^+)} \approx \frac{1-R}{1+R}
	=\frac{1}{2n_\t{m}+1}.
\end{equation}

Our objective is to increase the sideband asymmetry $1-R$ in the heterodyne spectrum, and thereby $\xi$,
by actively cold damping \cite{CouHeid01,WilSudKip15} the mechanical oscillator 
using the homodyne measurement as an error signal.
Concretely, the homodyne signal in the phase quadrature ($\theta=\pi/2$) is imprinted onto the
amplitude quadrature of an independent \textit{feedback} laser resonant with an auxiliary cavity mode 
($\lambda\approx 840\, \t{nm}$). 
The loop delay is tuned in order to produce a purely viscous radiation pressure feedback force, effectively
coupling the oscillator at a rate $\Gamma_\t{fb}\approx g_\t{fb} \Gamma_m$ to a cold bath with an occupation
equal to the phonon-equivalent homodyne imprecision $n_\t{imp}^\t{hom}=
\bar{S}_{x}^\t{imp,hom}(\Omega_\t{m})/2\bar{S}_{x}^\t{zp}(\Omega_m)$ (here $g_\t{fb}$ is the dimensionless
gain of the feedback loop).
The occupation of the oscillator is thereby reduced to,
\begin{equation}\label{eq:colddamping}
	n_\t{m}+\frac{1}{2}\approx 
		\frac{n_\t{tot}}{g_\t{fb}} +
		g_\t{fb} n_\t{imp}^\t{hom} \geq 2\sqrt{n_\t{tot}n_\t{imp}^\t{hom}},
\end{equation}
with the minimum achieved at an optimal gain of 
$g_\t{fb}^\t{opt}=\sqrt{n_\t{tot}/n_\t{imp}^\t{hom}}$. (Here, $n_\t{tot}=n_\t{th}+n_\t{ba}$ is the 
	effective bath occupation of the mechanical oscillator, including measurement back-action.)
Notably, cold-damping allows access to 
$n_\t{m}\rightarrow0$ when a highly efficient measurement is used, corresponding to an imprecision-back-action
product approaching the uncertainty limit $2\sqrt{n_\t{tot}n_\t{imp}^\t{hom}} \rightarrow \frac{1}{2}$.
Two regimes 
may be identified: (1) an \textit{efficient} feedback regime ($g_\t{fb}<g_\t{fb}^\t{opt}$), 
in which the motion of the oscillator -- resulting from the thermal noise and 
measurement back action -- is efficiently suppressed;
(2) an \textit{inefficient} feedback regime, in which
thermal force and measurement back-action are overwhelmed by \textit{feedback back-action}
$n_\t{fb}=g_\t{fb}^2 n_\t{imp}^\t{hom}$ (i.e. feedback of homodyne imprecision
noise), resulting in an increase of $n_\t{m}$.
We explore these regimes in two experiments.

An experimental demonstration of efficient feedback cooling, where feedback back-action is
weak ($n_\t{fb}<n_\t{tot}$), is shown in \fref{fig2}. 
Here $n_\t{tot}\approx 7\cdot 10^4$, corresponding to an effective bath temperature of $13\, \t{K}$
(arising partly due to quantum measurement back-action, $n_\t{ba}\approx 4\cdot 10^4$ \cite{WilSudKip15}).
From the perspective of the heterodyne measurement, the objective is to `distill' a motional
sideband asymmetry of one phonon out of $n_\t{tot}$.
This is made possible by a low
shot-noise-limited homodyne imprecision of
$n_\t{imp}^\t{hom} \approx 1.2\cdot 10^{-4}$ (see \fref{fig2}d for details).
To trace out the cooling curve in \fref{fig2}a, the feedback gain is
tuned electronically while keeping all other experimental parameters (such as mean optical power and
laser-cavity detuning) fixed. Sideband ratio $R$ is extracted from fitting a Lorentzian
to each heterodyne sideband and taking the ratio of the fitted areas. The phonon occupation $n_m$ is inferred
from $R$ as well as the area beneath the lower sideband.
In-loop (homodyne) and out-of-loop (heterodyne) noise spectra are shown in \fref{fig2}b. 
As a characteristic of the efficient feedback regime, 
the area under the left sideband decreases linearly with $g_\t{fb}$,
corresponding to $n_m \propto g_\t{fb}^{-1}$ (red circles in \fref{fig2}a).
As the optimal gain is approached, the in-loop spectrum is reduced to the imprecision noise 
floor (black trace in \fref{fig2}b).
This transition coincides with the `appearance' of a sideband asymmetry of $1-R \approx 12\%$ 
($\xi \approx 6\%$), corresponding to $n_\t{m}\approx 7.3$. 

\begin{figure*}[t!]
	\includegraphics[scale=0.73]{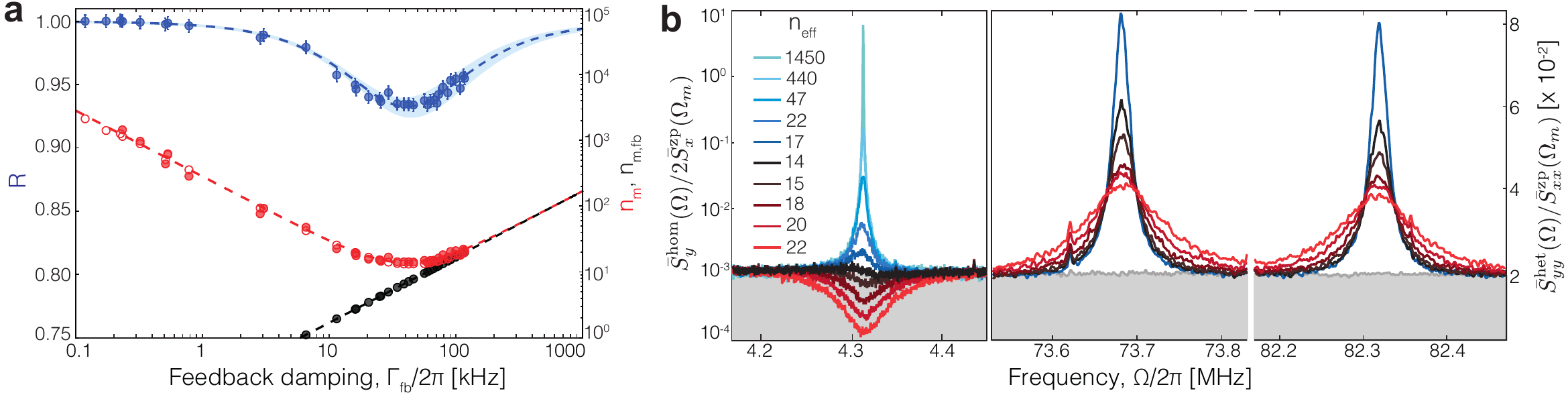}
	\caption{\label{fig3}
		\textbf{Appearance and disappearance of sideband asymmetry.} (a) Repeat of the experiment shown in 
		\fref{fig2}a with lower homodyne detection efficiency.   Feedback with the same range of gain results 
		in lower optimal asymmetry ($R\approx6\%$) and accesses to a `strong feedback' regime in which 
		feedback back-action ($n_\t{fb}$) dominates physical motion, resulting in reduced $R$.  
		Black points are an estimate of the mechanical occupation due to feedback back-action,
		$n_\t{m,fb}=\frac{\Gamma_m}{\Gamma_\t{fb}} n_\t{fb} = g_\t{fb}n_\t{imp}^\t{hom}$, 
		based on the noise floor of the homodyne spectra. (b) Left panel:
		In-loop homodyne spectra. In the strong feedback regime, noise is `squashed' (reduced below the 
		open-loop imprecision),
		corresponding to in-loop squeezing. Right panel: Out-of-loop heterodyne spectra.  
		Inefficient feedback manifests as an
		increase in the off-resonant noise power and reduced asymmetry.
	}
\end{figure*}

To confirm the faithfulness of these measurements, two major sources of error were investigated:

(1) Drift over the course of measurement can introduce small changes in the relative magnitude of 
$\bar{S}_{yy}^\t{het}(\Omega_\t{het}^\pm)$. In our experiment, this effect is mitigated by recording 
both heterodyne sidebands simultaneously.  Augmented by operating in the bad cavity regime
($\Omega_m/\kappa \sim 10^{-3}$),
and the exceptionally low imprecision of the heterodyne measurement, 
$n_\t{imp}^\t{het}=(4\eta_\t{het}C_0 n_c)^{-1} \approx 3\cdot 10^{-3}$ (see \fref{fig2}d),
statistical fluctuations of $R$ over the course of a typical measurement set can be as small as 0.5$\%$ 
(see \fref{fig2}c).
Error bars for $R$ in \fref{fig2}a are derived from the standard deviation of similar data sets, in addition 
to a small contribution
from the fit covariance matrix. At the largest damping rates, the reduced heterodyne signal-to-noise results in 
insufficient convergence of the periodogram estimate of the spectra (keeping 
acquisition time and analysis bandwidth fixed), leading to larger error bars, $\delta R =\pm 2\%$.

(2) Excess laser noise affects $R$
by producing additional imprecision-back-action correlations \cite{JayHarris12,Wein14}. Assuming a mean thermal photon 
occupation of $C_{qq(pp)}$ for the amplitude (phase) quadrature of the injected meter field, the correlator in 
\eqref{eq:Syy} becomes \cite{suppinfo},
\begin{equation}\label{eq:ScorrHet}
	\frac{2\re\bar{S}_{x_\t{ba}x_\t{imp}}^\t{het}(\Omega_\t{het}^{\pm})}{\bar{S}_{xx}^\t{zp}(\Omega_\t{m})} 
		= \mp\eta_\t{het}\left(\frac{1}{2}+C_{qq}\pm\frac{4\bar{\Delta}\Omega_m}{\kappa^2}C_{pp}\right),
\end{equation}
where $\eta_\t{het}$ is the heterodyne detection efficiency, and $\bar{\Delta}$ is the mean laser-cavity detuning. In 
our experiment, independent measurements reveal that 
$C_{qq}<0.01$ and $C_{pp}<30$ (owing partly to excess cavity frequency noise) 
for typical meter powers of $P_\t{in}<5\, 
\t{\mu W}$ \cite{suppinfo}. Operating on resonance $(\bar{\Delta} \approx 0)$ 
and in the bad-cavity regime substantially 
reduces sensitivity to $C_{pp}$. 
Using a typical value of $\bar{\Delta}= 0.01\cdot\kappa$, we estimate that $\frac{4\bar{\Delta} 
\Omega_m}{\kappa^2}C_{pp} < 0.005$ negligibly to \eqref{eq:ScorrHet}.

Having established that our measurements of motional sideband asymmetry are not contaminated by
classical artefacts, the results shown in \fref{fig2} may be interpreted 
as a `distillation' of quantum correlations using efficient feedback.
We now explore the complementary regime of inefficient feedback, where feedback back-action 
is stronger than 
the thermal force and measurement back-action ($n_\t{fb}>n_\t{tot}$).
We access this regime 
by changing the homodyne/heterodyne splitting ratio, thereby increasing the homodyne imprecision to
$n_\t{imp}^\t{hom}\approx 10^{-3}$.  
As shown in \fref{fig3}, increasing the gain 
beyond its optimum value (corresponding to $n_\t{m}\approx 13.4$ and $1-R\approx7\%$), results in a 
reduction of the homodyne signal 
below the shot-noise level (\fref{fig3}b left panel). 
Simultaneously, the areas of the heterodyne sidebands increase, while their asymmetry ($1-R$) decreases.
The discrepancy between `squashing' \cite{Taub95,Wise99} of the in-loop signal and 
the `disappearance' of sideband asymmetry relates to a basic difference between feedback back-action and meter 
back-action, namely, feedback back-action is correlated with the in-loop imprecision and not with
the out-of-loop imprecision \cite{Taub95}. 

Squashing of the in-loop signal is caused by
correlations between the feedback back-action driven motion $x_\t{fb}$ and the in-loop measurement 
imprecision \cite{suppinfo},
\begin{equation}\label{ScorrHom}
	\frac{2\t{Re} \bar{S}_{x_\t{fb}x_\t{imp}}^\t{hom}(\Omega_m)}{2\bar{S}_{xx}^\t{zp}(\Omega_m)} = 
		-n_\t{imp}^\t{hom}g_\t{fb}.
\end{equation}
represented by the negative-valued green trace in \fref{fig1}b (left panel). 
Interestingly, these classical correlations, in conjunction with the
generalized Heisenberg uncertainty principle \cite{BragKhal,Clerk10}
can be used to predict the transition from efficient to inefficient feedback; viz.
\begin{equation}\label{eq:generalizedUP}
	\bar{S}_{FF}\cdot \bar{S}_{xx}^\t{imp,hom} \geq 
		\frac{\hbar^2}{2}+(2\t{Re}\,\bar{S}_{Fx_\t{imp,hom}})^2,
\end{equation}
is saturated for $g_\t{fb}^\t{opt}=\sqrt{n_\t{tot}/n_\t{imp}^\t{hom}}$ (using $F_\t{fb}\propto g_\t{fb} 
x_\t{imp}^\t{hom}$ and \eqref{ScorrHom}).
The limits of feedback cooling, and the prospects for feedback-based enhancement of
quantum correlations, is related to the detection of meter fluctuations and the choice of feedback 
strategy -- optimization of either seems pertinent. 



\appendix
\section{Excess laser noise}\label{sec:noise}

The effect of laser noise on sideband asymmetry measurements is well-studied for cavity optomechanical
systems in the resolved sideband regime \cite{JayHarris12,AmirPain13}. 
In this case sidebands have been observed separately by scattering them
into the cavity with a probe laser red/blue detuned. 
Here we discuss the effect of laser noise on sideband asymmetry measurements in the 
``bad-cavity" regime ($\Omega_m \ll \kappa$), wherein a resonant probe is used to detect the sidebands 
simultaneously in a heterodyne measurement.  
A theoretical model is developed in
\secref{sec:noiseTheory}.  In \secref{sec:noiseExperiment}, we present measurements confirming the 
negligeable contribution of laser noise to the reported results.

\subsection{Contribution of excess noise for resonant probing and simultaneous detection of sidebands}
\label{sec:noiseTheory}

In our experiment, we probe the optomechanical system using a resonant laser at frequency $\omega_L$. 
The photon flux amplitude operator of the laser, $a_\t{in}(t)$, is assumed to have the form,
\begin{equation}
a_\t{in}(t) = e^{-i\omega_L t}(\bar{a}_\t{in}+\delta a_\t{in}(t)),
\end{equation}
where $\bar{a}_\t{in}=\sqrt{P_\t{in}/\hbar \omega_L}$ is the mean photon flux and the fluctuations 
$\delta a_\t{in}(t)$ satisfy,
\begin{equation}
[\delta a_\t{in}(t),\delta a_\t{in}^\dagger(t')] = \alpha\, \delta(t-t').
\end{equation}
Note that we explicitly ``tag'' the commutator so as to follow its contribution to the 
measured quantities \cite{Wein14}; in reality $\alpha =1$.

The canonically conjugate quadratures corresponding to the fluctuations are defined as
\begin{equation}
\begin{split}
	& \delta q_\t{in}(t) \eqdef \frac{\delta a_\t{in}(t) +\delta a_\t{in}^\dagger(t)}{\sqrt{2}},\\
	& \delta p_\t{in}(t) \eqdef \frac{\delta a_\t{in}(t)-\delta a_\t{in}^\dagger(t)}{i\sqrt{2}},
\end{split}
\end{equation}
so that 
\begin{equation}
[\delta q_\t{in}(t),\delta p_\t{in}(t')] = i\alpha\, \delta(t-t').
\end{equation}
Excess (``classical'') noise in the laser is modeled as Gaussian fluctuations, for which,
\begin{equation}\label{eq:noiseCorrelatorsFull}
\begin{split}
&\begin{pmatrix}
\avg{\delta q_\t{in}(t) \delta q_\t{in}(t')} & 
\avg{\delta q_\t{in}(t) \delta p_\t{in}(t')} \\
\avg{\delta p_\t{in}(t) \delta q_\t{in}(t')} & 
\avg{\delta p(t) \delta p(t')}
\end{pmatrix} \\
&\qquad= \frac{1}{2}\begin{pmatrix}
\alpha+2C_{qq} & i\alpha+ 2C_{qp} \\
-i\alpha+2C_{qp} & \alpha+2C_{pp}
\end{pmatrix} \delta (t-t').
\end{split}
\end{equation}
The terms $C_{ij}$ $(i=q,p)$ represent the noise in excess of
the fundamental vacuum fluctuations in the field quadratures, distributed uniformly 
(i.e. ``white'') in frequency. 
We henceforth omit the cross-correlation $C_{qp}$ and attempt to bound
its effect via an appropriate inequality
\footnote{in addition, it is known that for semiconductor lasers, phase-amplitude correlations
	are limited to frequencies close to their relaxation oscillation frequency \cite{Vah83,Exet92}; 
	the latter is typically at a few GHz from the carrier \cite{Kip13} -- irrelevant for our experiment}
(see \secref{sec:CqpInequality}). 
Thus,
\begin{equation}\label{eq:noiseCorrelators}
\begin{split}
&\begin{pmatrix}
\langle \delta a_\t{in}(t) \delta a_\t{in}(t')\rangle & 
\langle \delta a_\t{in}(t) \delta a_\t{in}^\dagger(t') \rangle \\
\langle \delta a_\t{in}^\dagger(t) \delta a_\t{in}(t') \rangle & 
\langle \delta a_\t{in}^\dagger (t) \delta a_\t{in}^\dagger(t') \rangle
\end{pmatrix} \\
&\qquad = 
\frac{1}{2}\begin{pmatrix}
C_{qq}-C_{pp} & 2\alpha +C_{qq}+C_{pp} \\
C_{qq}+C_{pp} & C_{qq}-C_{pp}
\end{pmatrix}.
\end{split}
\end{equation}

We now consider an optomechanical system where the optical cavity is driven by a noisy input field
satisfying \eqref{eq:noiseCorrelators}.
The mechanical oscillator couples to the cavity field via radiation pressure and is additionally driven by a 
thermal Langevin force.
Fluctuations of the intracavity field amplitude ($\delta a$) and the mechanical oscillator amplitude 
($\delta b$) around their stable steady states satisfy \cite{AspKipMar14}
\begin{eqnarray}
\dot{\delta a} & = & +i\Delta \delta a -\frac{\kappa}{2} \delta a 
+ig(\delta b+\delta b^{\dagger})+\sqrt{\kappa}\, \delta a_\t{in} \label{eq:langevind}\\
\dot{\delta b} & = & -i\Omega_{m} \delta b-\frac{\Gamma_{m}}{2}\delta b 
+i(g^\star\delta a+g\delta a^{\dagger})+\sqrt{\Gamma_{m}}\,\delta b_\t{in}. \label{eq:langevinc}
\end{eqnarray}
Here $\Delta = \omega_L-\omega_c$ is the laser detuning, $g = g_0 \bar{a}$ is the dressed (``multi-photon'') 
optomechanical coupling rate, and $\bar{a} = \frac{\sqrt{\kappa} \bar{a}_\textrm{in}}{\frac{\kappa}{2}-i\Delta}$ 
is the mean intracavity field amplitude.
We have also assumed here that the cavity decay rate is dominated by its external coupling, i.e .
$\kappa =\kappa_0 +\kappa_\t{ex} \approx \kappa_\t{ex}$.
The mechanical Langevin noise correlators are
\begin{align}
\langle \delta b_\t{in}(t) \delta b^\dagger_\t{in}(t') \rangle 
&= (n_\t{th} + \beta) \delta(t-t')\\
\langle \delta b^\dagger_\t{in}(t) \delta b_\t{in}(t') \rangle 
&= n_\t{th}\, \delta(t-t'),
\end{align}
where $n_\t{th}$ is the ambient mean thermal phonon occupation of the oscillator. Note that we also ``tag'' the
contribution due to the zero-point fluctuation of the thermal bath to determine its role in the observables; in
reality $\beta =1$.

Equations (\ref{eq:langevind}) and (\ref{eq:langevinc}) can be solved in the Fourier domain,
\begin{widetext}
\begin{align}\label{solution}
\delta a[\Omega] & = \chi_c[\Omega] \left[ \sqrt{\kappa}\,\delta a_\t{in}[\Omega]
+i g (\delta b[\Omega]+\delta b^\dagger[\Omega])\right]\\
\delta a^\dagger[\Omega] & =  \delta a[-\Omega]^\dagger =
\chi_c^\star[-\Omega] \left[\sqrt{\kappa}\,\delta a^\dagger_\t{in}[\Omega] 
-i g^\star (\delta b[\Omega]+\delta b^\dagger[\Omega])\right]\\ \nonumber 
\begin{pmatrix}
\delta b[\Omega] \\ \delta b^\dagger[\Omega]
\end{pmatrix} = &
\frac{\sqrt{\Gamma_m}}{\mathcal{N}[\Omega]}
\begin{pmatrix}
\chi^{\star -1}_m[-\Omega] - i \Sigma[\Omega] & -i\Sigma[\Omega]\\
+i\Sigma[\Omega] & \chi^{-1}_m[\Omega] + i\Sigma[\Omega]
\end{pmatrix}
\begin{pmatrix}
\delta b_\t{in}[\Omega] \\ \delta b^\dagger_\t{in}[\Omega]
\end{pmatrix} \\
& + \frac{i \sqrt{\kappa}}{\mathcal{N}[\Omega]}
\begin{pmatrix}
g^\star\chi_m^{\star -1}[-\Omega] \chi_c[\Omega] & g \chi_m^{\star -1}[-\Omega] \chi_c^\star[-\Omega]\\
-g^\star\chi_m^{-1}[\Omega] \chi_c[\Omega] & -g\chi_m^{-1}[\Omega] \chi_c^\star[-\Omega]
\end{pmatrix}
\begin{pmatrix}
\delta a_\t{in}[\Omega] \\ \delta a^\dagger_\t{in}[\Omega]
\end{pmatrix}. \nonumber
\end{align}
\end{widetext}
Here $\chi_m$ and $\chi_c$ are the bare mechanical and cavity response functions, respectively, given by,
\begin{equation}
\begin{split}
	\chi_{m}[\Omega] &\eqdef [\Gamma_{m}/2-i(\Omega-\Omega_{m})]^{-1}, \\
	\chi_{c}[\Omega] &\eqdef [\kappa/2-i(\Omega+\Delta)]^{-1}.
\end{split}
\end{equation}
$\Sigma[\Omega]$ is the mechanical ``self-energy'',
\begin{equation}\label{self-energy}
\Sigma[\Omega]=-i|g|^{2}(\chi_c[\Omega]-\chi_c^{\star}[-\Omega]) = \Sigma^\star[-\Omega],
\end{equation}
which describes the modification to the mechanical response due to radiation pressure, and
\begin{equation}
\mathcal{N}[\Omega]=\chi^{-1}_{m}[\Omega]\chi^{\star-1}_{m}[-\Omega] +2\Omega_m \Sigma[\Omega] = 
\mathcal{N}^\star[-\Omega].
\end{equation}

The input-output relation \cite{Gard85}, $\delta a_\t{out} = \delta a_\t{in} - \sqrt{\kappa} \, \delta a$, 
gives the fluctuations of the output fields in terms of the fluctuations of the input fields:
\begin{align}\label{aout}
\delta a_\t{out} = 
&A[\Omega]\delta a_\t{in} +B[\Omega]\delta a_\t{in}^\dagger+
C[\Omega]\delta b_\t{in}+D[\Omega]\delta b_\t{in}^{\dagger}
\end{align}
where,
\begin{widetext}
\begin{equation}\label{ABCD}
\begin{split}
A[\Omega] &= 
1 - \kappa \chi_c[\Omega] - \frac{2i|g|^2\kappa\Omega_m\chi_c[\Omega]^2}{\mathcal{N}[\Omega]} \approx
-\left(1+4i\frac{\Delta}{\kappa}\right)\left(1+C_0n_c\, 
\frac{2i\Omega_m\Gamma_m}{\mathcal{N}[\Omega]}\right)\\
B[\Omega] &= -\frac{2ig^2 \kappa \Omega_m\chi_c[\Omega]\chi_c^\star[-\Omega]}{\mathcal{N}[\Omega]} \approx
-C_0 n_c\, \frac{2i\Omega_m \Gamma_m}{\mathcal{N}[\Omega]} \\
C[\Omega] &= -\frac{i g \sqrt{\kappa\Gamma_m}}{\mathcal{N}[\Omega]} 
\chi_c[\Omega] \chi_m^{\star -1}[-\Omega] \approx
-i\sqrt{C_0 n_c}\, \left(1+2i\frac{\Delta}{\kappa}\right)\, \Gamma_m \chi_m[\Omega] \\
D[\Omega] &= -\frac{i g \sqrt{\kappa\Gamma_m}}{\mathcal{N}[\Omega]} 
\chi_c[\Omega] \chi_m^{-1}[\Omega] \approx
-i\sqrt{C_0 n_c}\, \left(1+2i\frac{\Delta}{\kappa}\right)\, \Gamma_m \chi_m^\star[-\Omega].
\end{split}
\end{equation}
\end{widetext}
Here approximate expressions are given for the case of interest, namely, resonant probing ($\abs{\Delta} \ll \kappa$), 
small sideband resolution ($\Omega_m \ll \kappa$), and weak coupling ($\abs{g} \ll \kappa$). 
We have also introduced the single-photon cooperativity, 
$C_0 = 4g_0^2/(\kappa \Gamma_m) $, and the mean intracavity photon number, $n_c =\abs{\bar{a}}^2$.

Balanced heterodyne detection of the cavity output is used to measure motional sideband asymmetry.
The output field is superposed on a balanced beamsplitter with a frequency-shifted
local oscillator,
\begin{equation}
a_\t{LO} = e^{-i (\omega_\t{L}+\Omega_\t{IF}) t}(\bar{a}_\t{LO} + \delta a_\t{LO}).
\end{equation}
The fields at the output of the beamsplitter, $\tfrac{1}{\sqrt{2}}(a_\t{LO} \pm 
a_\t{out})$, are detected with identical square-law detectors, whose photocurrents are subtracted.
Note the implicit assumption that the local oscillator and signal paths are balanced in length; together
with a balance of power beyond the combining beamsplitter, this ensures suppression of common-mode
excess noise \cite{Shap85}.

The difference photocurrent is described by the operator,
\begin{equation}
I \propto a_\text{LO}^\dagger a_\text{out} + \text{H.c.}.
\end{equation}
When $\bar{a}_\t{LO} \gg \bar{a}_\t{out}$, fluctuations in the photocurrent are described by
\begin{equation}
\delta I(t) \propto e^{-i\Omega_\t{IF}t} \bar{a}_\t{LO}^\star\, \delta a_\t{out}(t) + \t{H.c.}.
\end{equation}
The power spectrum of the heterodyne photocurrent is proportional to
\begin{equation}\label{SiHet}
\bar{S}_{II}^\t{het}(\Omega) = \frac{1}{2} \int_{-\infty}^{\infty} \avg{\overline{\left\{ \delta I(t+t'),
		\delta I(t')\right\}}} e^{i\Omega t}dt,
\end{equation}
where we have introduced the (time-averaged) current correlator,
\begin{equation}\label{corrIhet}
\begin{split}
\overline{\left\{ \delta I(t+t'),\delta I(t')\right\}} &\propto 
e^{-i \Omega_\text{IF} t} 
\left\{ \delta a_\t{out}^\dagger(t),\delta a_\t{out}(0) \right\} \\ 
&\qquad+e^{+ i \Omega_\text{IF} t} 
\left\{ \delta a_\t{out}(t),\delta a_\t{out}^\dagger(0) \right\}.
\end{split}
\end{equation}

Assuming $\Omega_\text{IF} \gg \Omega_m >0$, we obtain for the balanced heterodyne spectrum normalized to the 
local oscillator shot noise,
\begin{widetext}
\begin{align}\label{eq:spectrumHeterodyne}
\bar{S}_{II}^\t{het}(\Omega-\Omega_\t{IF}) \approx \alpha + 4C_0 n_c 
& \left[
\tfrac{\Gamma_m^2}{4} \abs{\chi_m[-\Omega]}^2 \left(n_\t{tot}+\tfrac{\beta}{2}
-\left(\tfrac{\alpha}{2}+C_{qq}\right) 
+\tfrac{4\Delta \Omega_m}{\kappa^2}C_{pp}
\right) \right. \\ \nonumber
& \left. + \tfrac{\Gamma_m^2}{4} \abs{\chi_m[\Omega]}^2 \left(n_\t{tot}+\tfrac{\beta}{2}
+\left(\tfrac{\alpha}{2}+C_{qq}\right)+\tfrac{4\Delta \Omega_m}{\kappa^2}C_{pp} \right)
\right].
\end{align}
\end{widetext}
This represents the heterodyne spectrum measured in the experiment and depicted in Fig.2 and Fig.3 of the 
main text. Here the total bath occupation, arising from the ambient thermal bath and
the measurement back-action due to the meter beam, is given by,
\begin{equation}
n_\t{tot} = n_\t{th} + \underbrace{C_0 n_c \left(\tfrac{\alpha}{2}+C_{qq}+
	\left(\tfrac{4\Delta \Omega_m}{\kappa^2}\right)^2 C_{pp}\right)}_{n_\t{ba}}.
\end{equation}
The \textit{sideband ratio} extracted from such a spectrum is,
\begin{equation}\label{eq:R}
\begin{split} 
R & \eqdef \frac{\int_{0^+}^{+\infty} 
	(\bar{S}_{II}^\t{het}(\Omega-\Omega_\t{IF})-\bar{S}_{II}^\t{het}(\Omega=\Omega_\t{IF}^+)) \frac{d\Omega}{2\pi} }
{\int_{-\infty}^{0^{-}} 
	(\bar{S}_{II}^\t{het}(\Omega-\Omega_\t{IF})-\bar{S}_{II}^\t{het}(\Omega=\Omega_\t{IF}^-)) \frac{d\Omega}{2\pi} 
}\\ &=
\frac{n_\t{tot}+\tfrac{\beta-\alpha}{2}-C_{qq}+\tfrac{4\Delta\Omega_m}{\kappa^2}C_{pp}}
{n_\t{tot}+\tfrac{\beta+\alpha}{2}+C_{qq}+\tfrac{4\Delta \Omega_m}{\kappa^2}C_{pp}}\\ &=
\frac{n_\t{tot}+\left(\tfrac{4\Delta \Omega_m}{\kappa^2}C_{pp}-C_{qq}\right)}
{n_\t{tot}+1+\left(\tfrac{4\Delta \Omega_m}{\kappa^2}C_{pp}+C_{qq}\right)}.
\end{split} 
\end{equation}

Firstly, characteristic of linear detection, deviation of
$R$ from unity in the ideal case $(C_{qq}=0=C_{pp})$ is due to correlations developed between the quantum-back-action
driven mechanical motion and the detection process \cite{AmirPain13,Wein14}. 
When $C_{qq}$ and $C_{pp}$ are finite, classical correlations are established that affect $R$.
The response of the cavity (for $\Delta/\kappa \approx 0$) ensures that excess classical correlations due to
input amplitude noise lead to an enhanced asymmetry, whereas
those arising from input phase noise lead to a common increase in the sideband noise power.

\subsubsection{Expression for $\bar{S}_{yy}^\t{het}(\Omega)$}

In order to compare with Eq.(1) of the main text, we identify the heterodyne spectrum \eqref{eq:spectrumHeterodyne}
with that of a position-equivalent heterodyne observable $y_\t{het}$, viz.,
\begin{widetext}
\begin{equation}
\begin{split} 
\bar{S}_{yy}^\t{het}(\Omega-\Omega_\t{IF}) = 
& \underbrace{ 
	\left(\frac{1}{4C_0 n_c}\right)\bar{S}_{xx}^\t{zp}(\Omega_m) 
}_{\bar{S}_{xx}^\t{imp,het}(\Omega)} \\
& + \underbrace{
	\tfrac{\Gamma_m^2}{4}\left(\abs{\chi_m[-\Omega]}^2 + 
	\abs{\chi_m[\Omega]}^2 \right)\left(n_\t{tot}+\tfrac{1}{2}\right)\bar{S}_{xx}^\t{zp}(\Omega_m)
}_{\bar{S}_{xx}^\t{tot}(\Omega)}\\
& + \underbrace{
	\tfrac{\Gamma_m^2}{4}\left(
	\abs{\chi_m[-\Omega]}^2 
	\left(\tfrac{1}{2}+C_{qq}+\tfrac{4\Delta \Omega_m}{\kappa^2}C_{pp}\right)
	+ \abs{\chi_m[\Omega]}^2 
	\left(-\tfrac{1}{2}-C_{qq}+\tfrac{4\Delta \Omega_m}{\kappa^2}C_{pp}\right) \right)
}_{2\t{Re}\, \bar{S}_{x_\t{ba}x_\t{imp}}^\t{het}(\Omega)}.
\end{split}
\end{equation}
\end{widetext}
The identification is made by comparing the magnitude of the 
total thermal noise signal $\bar{S}_{xx}^\t{tot}$.

\subsubsection{Sensitivity of heterodyne and homodyne readout}

In the main text, frequent use is made of the phonon-equivalent sensitivity of the heterodyne and homodyne detectors.
The sensitivity of balanced heterodyne detection (for the ideal case $\eta_\t{het}=1$), quantified as 
imprecision quanta,
\begin{equation}
n_\t{imp}^\t{het} = (4\eta_\t{het}C_0 n_c)^{-1}
\end{equation}
is reduced by a factor of 4 compared to balanced homodyne detection (for the ideal case $\eta_\t{hom}=1$) of the 
phase quadrature of the output field, for which
\begin{equation}
n_\t{imp}^\t{hom} = (16\eta_\t{hom}C_0 n_c)^{-1}.
\end{equation}
This loss arises in equal part due to (a) the fact that the heterodyne spectrum is double-sided, and, (b) the detection 
of both quadratures of the output field. 

\subsection{Measurement of excess laser noise}
\label{sec:noiseExperiment}

\begin{figure*}
	\centering
	\includegraphics[scale=0.85]{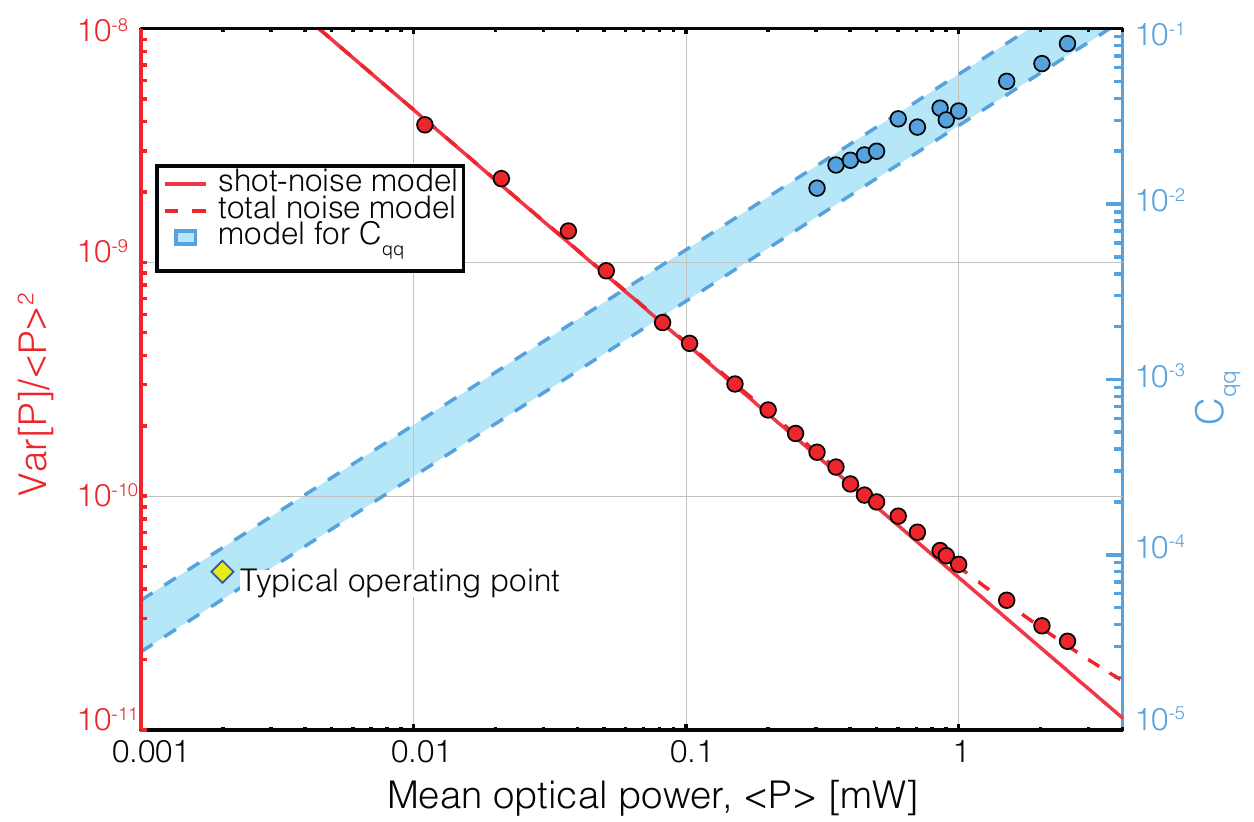}
	\caption{\label{fig:Cqq} Integrated (in a $100\, \t{kHz}$ band) relative intensity noise 
		$\frac{\t{Var}[P]}{\avg{P}^2}\eqdef\int \bar{S}_\t{RIN}(\Omega\approx \Omega_m) \frac{d\Omega}{2\pi}$ 
		versus mean optical power. 
		Deviation from shot-noise scaling is evident for $\avg{P}\gtrsim 1\, \t{mW}$, attributed to
		classical amplitude noise.}
\end{figure*}

\subsubsection{Excess amplitude noise}

In order to measure the noise in the amplitude quadrature, we employ direct photodetection of the probe laser. The
measurement is made at the output of the tapered fiber, with the fiber retracted from the cavity.
Analysis of the resulting photocurrent reveals the single-sided spectrum of the incident 
optical intensity (referred here for convenience to the incident optical power $P = \hbar \omega_L \dot{n}$),
\begin{equation}
\bar{S}_P(\Omega) 
= (\hbar \omega_L)^2\cdot 2\bar{S}_{\dot{n}\dot{n}}(\Omega) 
= (\hbar \omega_L)^2\cdot 2\avg{\dot{n}}(1+2C_{qq}).
\end{equation}

A convenient characterization of the intensity noise is via the relative intensity noise (RIN) spectrum,
\begin{equation}
\bar{S}_\t{RIN}(\Omega) \eqdef \frac{\bar{S}_P(\Omega)}{\avg{P}^2} 
\end{equation}
excess amplitude noise manifests as a deviation from the shot-noise scaling 
$\propto \frac{1}{\avg{P}}$; more precisely,
\begin{equation}\label{eq:CqqRIN}
C_{qq} = \frac{1}{2}\left(\frac{\avg{\dot{n}}}{2}\, \bar{S}_\t{RIN}(\Omega) -1\right).
\end{equation}
\fref{fig:Cqq} shows an inference of $C_{qq}$ using \eqref{eq:CqqRIN} and a measurement of
$\bar{S}_\t{RIN}(\Omega)$ versus mean optical power. For typical 
experimental conditions ($\avg{P}=1-5\, \t{\mu W}$), $C_{qq} \ll 0.01 $, so that its contribution to sideband asymmetry
is negligible.

\subsubsection{Excess phase noise}

Noise in the phase quadrature of the field leaking from the cavity is measured using balanced homodyne detection.
This signal reveals phase noise originating from the input laser as well as apparent phase noise from the cavity. 
Referred to cavity frequency noise, the homodyne photocurrent spectral density is given by,
\begin{equation}
\begin{split}
\bar{S}_{\omega}(\Omega) &= \Omega^2\bar{S}_{\phi}(\Omega) = 
\Omega^2\left( \bar{S}_{\phi}^\t{in,shot}(\Omega) +\bar{S}_{\phi}^\t{in,ex}(\Omega) \right. \\
&\qquad \left. +\bar{S}_{\phi}^\t{cav,ex}(\Omega) +\bar{S}_{\phi}^\t{cav,mech}(\Omega) \right).
\end{split}
\end{equation}
$\bar{S}_\omega$ contains contributions from laser phase noise (shot and excess), cavity substrate 
noise (including thermorefractive \cite{Gor04} and thermomechanical noise \cite{Gill93}) 
and thermal motion of other modes 
of the mechanical resonator.
The total excess noise in the phase quadrature is modeled by $C_{pp}$, which allows us to infer
the latter using,
\begin{equation}\label{eq:CppSomega}
\frac{C_{pp}}{\avg{\dot{n}}} = \bar{S}_\phi^\t{in,ex}(\Omega_m) +\bar{S}_\phi^\t{cav,ex}(\Omega_m) .
\end{equation}

\begin{figure*}
	\centering
	\includegraphics[scale=0.75]{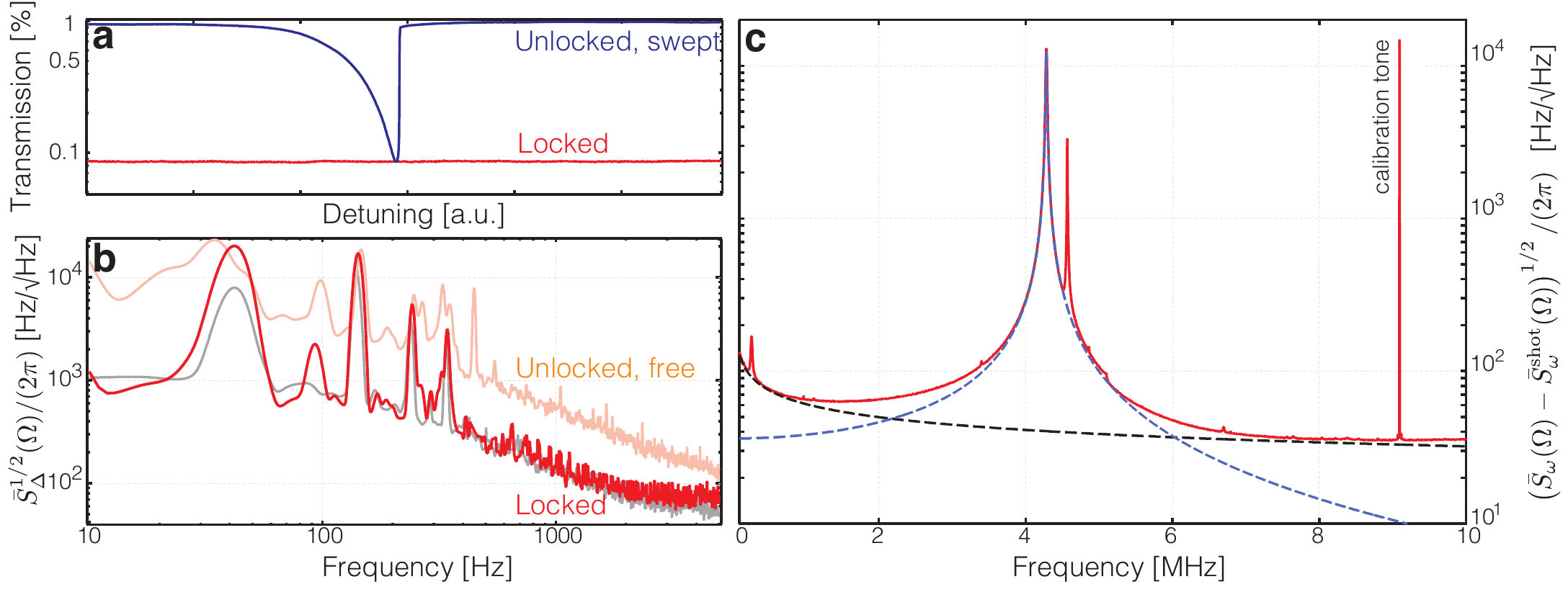}
	\caption{\label{fig:Cpp} (a) Residual detuning offset at DC estimated from transmission signal
		when the laser is locked to cavity. 
		(b) Spectrum analysis of the lock error signal, generated via frequency-modulation spectroscopy 
		\cite{Bjork80}, reveals low
		frequency detuning jitter; when locked (red), apparent detuning noise is limited by electronic noise (gray)
		in the feedback loop, predominantly from the photodetector. 
		(c) Excess frequency noise around the mechanical frequency inferred from a
		balanced homodyne measurement of the cavity output on resonance. The shot-noise-subtracted signal (red)
		is composed of the thermomechanical motion of the mechanical mode (blue dashed) and a contribution from
		excess frequency noise in the laser and cavity substrate (black dashed).}
\end{figure*}

\fref{fig:Cpp}c shows a homodyne measurement made with $3\, \t{mW}$ of local oscillator power, whose shot-noise has
been subtracted. The spectrum is calibrated by referencing it against a known phase modulation tone injected at the 
input of the homodyne interferometer \cite{GorKip10}. The total excess frequency noise (red) is dominated by
thermal motion of the in-plane and out-of-plane modes, both of which are gas damped for this measurement. 
A joint fit to (a) a model of a velocity-damped oscillator (blue, dashed) and, (b) a model combining 
thermorefractive \cite{BragGor00,Gor04} and white frequency noise (black, dashed), gives an estimate 
of $\bar{S}_{\omega}^\t{ex}(\Omega)$. Frequency noise intrinsic to the diode laser was independently
measured using an imbalanced interferometer, consistent with the model used to fit the total observed
frequency noise. Near the mechanical frequency, 
$\bar{S}_\omega^\t{ex}(\Omega_m)\approx 2\pi\cdot (35\, \t{Hz}/\sqrt{\t{Hz}})^2$, implying 
(via \eqref{eq:CppSomega}), $C_{pp}\approx 30$ (using signal power of $\approx 100\, \t{nW}$).

From this estimate of $C_{pp}$ we are able to bound two quantities. First, in conjunction with $C_{qq} \ll 0.01$, the 
excess noise cross-correlation is bounded as $C_{qp} \ll 1$. Secondly, referring to \eqref{eq:spectrumHeterodyne}, we are
able to estimate the contribution of phase noise to the heterodyne sideband. 
This contribution, characterized as an 
equivalent phonon occupation (since it adds positive noise power to either sideband),
\begin{equation}\label{eq:nphi}
n_\phi = \frac{\Delta}{\kappa}\frac{4\Omega_m}{\kappa}C_{pp},
\end{equation}
has a mean value determined by the mean offset in the detuning $\bar{\Delta}$. \fref{fig:Cpp}a allows an
estimate, $\bar{\Delta} \approx 0.01\cdot \kappa$, giving,
\begin{equation}\label{eq:nphiEquality}
\begin{split}
\bar{n}_\phi &= \frac{\bar{\Delta}}{\kappa}\frac{4\Omega_m}{\kappa}C_{pp}\\ 
&= 0.0052\cdot 
\left(\frac{\bar{\Delta}/\kappa}{0.01}\right)
4\left(\frac{\Omega_m/2\pi}{4.3\,\t{MHz}}\right)\left(\frac{1\,\t{GHz}}{\kappa/2\pi}\right)
\left(\frac{C_{pp}}{30}\right).
\end{split}
\end{equation}
Low frequency detuning noise $\delta \Delta$ (\fref{fig:Cpp}b) causes deviations from this mean, which are significant 
if their effect
is comparable to $\bar{n}_\phi$. We bound the probability for such ``large'' statistical excursions
using Chebyshev's inequality \cite{Cram},
\begin{equation}
\begin{split}
\t{Pr}(\abs{n_\phi-\bar{n}_\phi}>\bar{n}_\phi) 
&\leq \frac{\t{Var}[n_\phi]}{\bar{n}_\phi^2} \\
& = \left(\frac{4\Omega_m}{\kappa}\frac{C_{pp}}{\bar{n}_\phi}\right)^2 \frac{\t{Var}[\delta \Delta]}{\kappa^2} \\
& \approx 10^{-6}.
\end{split}
\end{equation}

We thus estimate that mean residual detuning is the leading contribution to phase noise contamination; the 
contamination, characterized as a phonon-equivalent noise power $\bar{n}_\phi = 0.005$ is however an 
insignificant contribution to the sideband ratio \eqref{eq:R}. 

Together with the bounds, $C_{qq}\ll 0.01$ and $C_{qp}\ll 1$, this implies that sources of classical noise may be 
excluded in the interpretation of the experimental data.

\subsection{Bounding the value of the classical noise cross-correlation $C_{qp}$}
\label{sec:CqpInequality}

In \cite{JayHarris12}, excess classical noise in the laser is modelled as an independent classical stochastic
process introduced explicitly into $\delta a_\t{in}$. The added term, being a classical stochastic process, obeys
a Cauchy-Schwarz inequality for its second moments, resulting in the inequality $C_{qp}\leq \sqrt{C_{qq}C_{pp}}$, which
may be employed to bound the magnitude of $C_{qp}$, given measurements of $C_{qq}$ and $C_{pp}$. 

Here we consider a more natural alternative, where the ansatz \eqref{eq:noiseCorrelators} is supposed to arise
from a choice of the underlying quantum state that models the classical component of the noise. From this
perspective, the ansatz in \eqref{eq:noiseCorrelators} is a valid one as long as it arises from a legitimate
quantum state $\rho$. The sufficient condition for the matrix in \eqref{eq:noiseCorrelators} to be
a valid covariance matrix is \cite{SimMuk94},
\begin{equation}
V\eqdef \begin{pmatrix}
\tfrac{1}{2}+C_{qq} & C_{qp} \\
C_{qp} & \tfrac{1}{2}+C_{pp}
\end{pmatrix} \geq 0.
\end{equation}
In particular, this implies that $\tr\, V \geq 0$ and $\t{det}\, V \geq 0$; the latter condition gives,
\begin{equation}\label{eq:CqpInequality}
\begin{split}
C_{qp}^2 & \leq C_{qq}C_{pp}+\frac{1}{2}(C_{qq}+C_{pp}) \\
& \leq C_{qq}C_{pp}+\sqrt{C_{qq}C_{pp}} \\
& = C_{qq}C_{pp}\left(1+\frac{1}{\sqrt{C_{qq}C_{pp}}}\right).
\end{split}
\end{equation}
Here, the second line is obtained by employing the inequality $C_{qq}+C_{pp}\geq 2\sqrt{C_{qq}C_{pp}}$ that
follows generally from the fact that $C_{qq,pp}$ are positive.

Ultimately, in the limit $C_{qq}C_{pp}\gg 1$, we recover the result in \cite{JayHarris12}, namely,
$C_{qp}\leq (C_{qq}C_{pp})^{1/2}$; however, in the opposite limit, $C_{qq}C_{pp}\ll 1$, 
the appropriate bound is $C_{qp}\leq (C_{qq}C_{pp})^{1/4}$,
and so employing the
conventional Cauchy-Schwartz inequality would lead to an under estimate of $C_{qp}$.

In our case, $C_{qq}C_{pp}\ll 0.3$, and \eqref{eq:CqpInequality} suggests $C_{qp}\ll 1$.

\section{Squeezing in homodyne detection}

In the experimentally relevant bad-cavity regime, $\Omega_m \ll \kappa$, resonant probing $\Delta = 0$, and 
quantum-noise limited probe laser, a significantly simplified analysis illustrates the presence of
correlations in the transmitted field. 

Following from \eqref{aout} and \eqref{ABCD}, the cavity transmission is given by,
\begin{equation}\label{aoutSq}
\delta a_\t{out}[\Omega] \approx -\delta a_\t{in}[\Omega] 
-i\frac{\sqrt{C_0 n_c \Gamma_m}}{x_\t{zp}}\left( x_\t{th}[\Omega]+x_\t{ba}[\Omega] \right),
\end{equation}
where the total motion, $x\eqdef x_\t{zp}(b+b^\dagger)$, has been partitioned into the (intrinsic) thermal
motion $x_\t{th}$ due to the ambient environment, 
\begin{equation}
\delta x_\t{th}[\Omega] \eqdef x_\t{zp}\sqrt{\Gamma_m}\left(\chi_m[\Omega]\delta b_\t{in}[\Omega] 
+\chi_m[-\Omega]^\ast \delta b_\t{in}^\dagger[\Omega]\right)
\end{equation}
and $x_\t{ba}$, the back-action driven motion,
\begin{equation}\label{xba}
\begin{split}
\delta x_\t{ba}[\Omega] &\eqdef x_\t{zp}\sqrt{2C_0 n_c \Gamma_m}\frac{2\Omega_m}{\mathcal{N}[\Omega]} \delta q_\t{in}[\Omega] \\
&\approx x_\t{zp}\sqrt{2C_0 n_c \Gamma_m} \frac{\delta q_\t{in}[\Omega]}{(\Omega-\Omega_m)-i(\Gamma_m/2)}
\end{split}
\end{equation}
due to the vacuum fluctuations in the amplitude quadrature of the input optical field. Note that the second
equality neglects dynamical back-action and assumes a high-Q mechanical oscillator.

In terms of the amplitude ($\delta q$) and phase ($\delta p$) quadratures, \eqref{aoutSq} takes the form,
\begin{widetext}
\begin{equation}
\begin{split}
\delta q_\t{out}[\Omega] &= -\delta q_\t{in}[\Omega] \\
\delta p_\t{out}[\Omega] &= -\delta p_\t{in}[\Omega] -\frac{\sqrt{2C_0 n_c \Gamma_m}}{x_\t{zp}}
\left(x_\t{th}[\Omega]+x_\t{ba}[\Omega]\right) \\
& = -\delta p_\t{in}[\Omega] 
-\sqrt{2C_0 n_c \Gamma_m} \frac{x_\t{th}[\Omega]}{x_\t{zp}} 
-\frac{2C_0 n_c \Gamma_m}{(\Omega-\Omega_m)-i(\Gamma_m/2)} \delta q_\t{in}[\Omega].
\end{split}
\end{equation}
\end{widetext}
Note that the transmitted phase quadrature has a component proportional to the transmitted amplitude quadrature,
leading to phase-amplitude correlations described by the (un-symmetrized, double-sided) cross-correlation spectrum,
\begin{equation}\label{SpqOut}
S_{pq}^\t{out}(\Omega) = -\frac{i}{2}+ \frac{C_0 n_c \Gamma_m}{(\Omega-\Omega_m)-i(\Gamma_m/2)}.
\end{equation}
where the first term is due to the commutation relation of the transmitted fields, while the second arises from 
correlations induced by the optomechanical interaction.

Homodyne detection of the phase quadrature, corresponding to a measurement of $\delta p_\t{out}$ alone, does not
give access to these optomechanically induced correlations. However, homodyne detection at a finite phase offset
$\theta$, corresponding to a measurement of,
\begin{equation}
\delta q_\t{out}^\theta [\Omega] \eqdef \delta q_\t{out}[\Omega] \cos \theta + \delta p_\t{out}[\Omega]\sin \theta,
\end{equation}
can directly access amplitude-phase correlations. Indeed, the homodyne photocurrent spectrum, 
$\bar{S}_{II}^{\t{hom},\theta} (\Omega) \propto \bar{S}_{qq}^\t{out,\theta}(\Omega)$, takes the form,
\begin{equation}\label{SiHomTheta}
\begin{split}	
\bar{S}_{II}^{\t{hom},\theta} (\Omega) &\propto \cos^2 \theta\, \bar{S}_{qq}^\t{out}(\Omega) 
+\sin^2 \theta\, \bar{S}_{pp}^\t{out}(\Omega)\\ 
&\qquad+ \sin(2\theta)\, \t{Re}\, S_{pq}^\t{out}(\Omega),
\end{split}	
\end{equation}
so that for $\theta \neq 0,\pi/2$, the correlation term is manifest.
Including the effect of non-ideal detection efficiency, $\eta_\t{hom}\leq 1$, and normalizing to shot-noise, 
the homodyne photocurrent spectrum is,
\begin{widetext}
\begin{equation}
\bar{S}_{II}^\t{hom,\theta}(\Omega) = 1 + 4C_0 n_c\eta_\t{hom} \frac{\bar{S}_{xx}(\Omega)}{x_\t{zp}^2} \sin^2 \theta
+ \underbrace{2C_0 n_c \eta_\t{hom}\frac{\Gamma_m(\Omega-\Omega_m)}{(\Omega-\Omega_m)^2+(\Gamma_m/2)^2} \sin(2\theta) 
}_{2\t{Re}\, \bar{S}_{x_\t{ba}x_\t{imp}}^\t{hom,\theta}(\Omega)}.
\end{equation}
\end{widetext}
The last term, anti-symmetric in frequency about the mechanical resonance frequency $\Omega_m$, can contribute
negatively to the photocurrent spectrum, leading to squeezing below the shot-noise level. The last term may be can be
identified as being due to correlations between the back-action driven motion $x_\t{ba}$, and the fluctuations of
the transmitted field that set the imprecision in homodyne detection. The above equation is used to fit the squeezing
spectrum in Fig. 2 of the main manuscript.

 \subsection{Relation to heterodyne sideband asymmetry}

 Following the discussion of heterodyne detection in \secref{sec:noiseTheory}, leading up to equations (\ref{SiHet}) and
 (\ref{corrIhet}), the heterodyne photocurrent spectrum centred around the intermediate frequency $\Omega_\t{IF}$ is
 given by,
 \begin{equation}
 \begin{split}
 \bar{S}_{II}^\t{het}(\Omega-\Omega_\t{IF}) &\propto \bar{S}_{qq}^\t{out}(\Omega)+\bar{S}_{pp}^\t{out}(\Omega) \\
	 &\quad+\t{Im}\left(S_{qp}^\t{out}(-\Omega)-S_{pq}^\t{out}(+\Omega) \right),
 \end{split}
 \end{equation}
 where $\Omega \geq 0$ and the approximation $\Omega_\t{IF} \gg \Omega_m \gg 0$ is used. 
 Indeed, the asymmetry in the heterodyne spectrum, about $\Omega=\Omega_\t{IF}$, arises from the imaginary part of
 the quantum correlations between the phase and amplitude of the transmitted field.
 Compared to the analogous expression for the homodyne photocurrent spectrum in \eqref{SiHomTheta}, where the real part
 of the correlation leads to optical squeezing, it is the imaginary part of the phase-amplitude
 correlation (\eqref{SpqOut}) that contributes to sideband asymmetry.

\section{Displacement spectrum of a cold-damped mechanical oscillator}

Here we recall a few useful expressions for the displacement spectrum of a cold-damped mechanical 
oscillator \cite{CouHeid01,WilSudKip15}. We denote by $x$ the physical displacement
of the oscillator, and by $y_\t{hom}=x+x_\t{imp}^\t{hom}$, the apparent displacement measured at the in-loop (homodyne)
detector. Following the arguments detailed in the supplementary information of \cite{WilSudKip15}, we get,
\begin{equation}\label{eq:SxSxmeas}
\begin{split}
\bar{S}_{x}(\Omega) &= 
\underbrace{\abs{\chi_\t{eff}(\Omega)}^2 (2n_\t{tot}+1)\bar{S}_{x}^\t{zp}(\Omega_\t{m})}_{
	\bar{S}_x^\t{tot}(\Omega)} \\ 
&+
\underbrace{\abs{\chi_\t{eff}(\Omega)}^2 (2n_\t{imp}^\t{hom}g_\t{fb}^2)\bar{S}_{x}^\t{zp}(\Omega_\t{m})}_{
	\bar{S}_x^\t{fb}(\Omega)}
\end{split}
\end{equation}
for the physical displacement spectrum, and,
\begin{equation}
\begin{split}
\bar{S}_{y}^\t{hom}(\Omega) &= 
\underbrace{2n_\t{imp}^\t{hom}\bar{S}_{x}^\t{zp}(\Omega_\t{m})}_{
	\bar{S}_x^\t{imp,hom}(\Omega)} \\
&+
\underbrace{\abs{\chi_\t{eff}(\Omega)}^2 (2n_\t{tot}+1)\bar{S}_{x}^\t{zp}(\Omega_\t{m})}_{
	\bar{S}_x^\t{tot}(\Omega)} \\
&+
\underbrace{\abs{\chi_\t{eff}(\Omega)}^2 (-2n_\t{imp}^\t{hom} g_\t{fb})\bar{S}_{x}^\t{zp}(\Omega_\t{m})}_{
	2\t{Re}\bar{S}_{x_\t{fb}x_\t{imp}}^\t{hom}(\Omega)}
\end{split}
\end{equation}
for the apparent displacement spectrum. Here, the effective susceptibility for the displacement is given by,
\begin{equation}
\chi_\t{eff} = \frac{\Omega_m \Gamma_m}{(\Omega_m^2 -\Omega^2)+i\Omega (\Gamma_m+\Gamma_\t{fb})},
\end{equation}
where $\Gamma_\t{fb}=\Gamma_m g_\t{fb}$ is the feedback damping rate. In the main text we use the approximation
$\Gamma_m +\Gamma_\t{fb}\approx \Gamma_\t{fb}$.


\begin{thebibliography}{41}%
	\makeatletter
	\providecommand \@ifxundefined [1]{%
		\@ifx{#1\undefined}
	}%
	\providecommand \@ifnum [1]{%
		\ifnum #1\expandafter \@firstoftwo
		\else \expandafter \@secondoftwo
		\fi
	}%
	\providecommand \@ifx [1]{%
		\ifx #1\expandafter \@firstoftwo
		\else \expandafter \@secondoftwo
		\fi
	}%
	\providecommand \natexlab [1]{#1}%
	\providecommand \enquote  [1]{``#1''}%
	\providecommand \bibnamefont  [1]{#1}%
	\providecommand \bibfnamefont [1]{#1}%
	\providecommand \citenamefont [1]{#1}%
	\providecommand \href@noop [0]{\@secondoftwo}%
	\providecommand \href [0]{\begingroup \@sanitize@url \@href}%
	\providecommand \@href[1]{\@@startlink{#1}\@@href}%
	\providecommand \@@href[1]{\endgroup#1\@@endlink}%
	\providecommand \@sanitize@url [0]{\catcode `\\12\catcode `\$12\catcode
		`\&12\catcode `\#12\catcode `\^12\catcode `\_12\catcode `\%12\relax}%
	\providecommand \@@startlink[1]{}%
	\providecommand \@@endlink[0]{}%
	\providecommand \url  [0]{\begingroup\@sanitize@url \@url }%
	\providecommand \@url [1]{\endgroup\@href {#1}{\urlprefix }}%
	\providecommand \urlprefix  [0]{URL }%
	\providecommand \Eprint [0]{\href }%
	\providecommand \doibase [0]{http://dx.doi.org/}%
	\providecommand \selectlanguage [0]{\@gobble}%
	\providecommand \bibinfo  [0]{\@secondoftwo}%
	\providecommand \bibfield  [0]{\@secondoftwo}%
	\providecommand \translation [1]{[#1]}%
	\providecommand \BibitemOpen [0]{}%
	\providecommand \bibitemStop [0]{}%
	\providecommand \bibitemNoStop [0]{.\EOS\space}%
	\providecommand \EOS [0]{\spacefactor3000\relax}%
	\providecommand \BibitemShut  [1]{\csname bibitem#1\endcsname}%
	\let\auto@bib@innerbib\@empty
	\bibitem [{\citenamefont {Wiseman}\ and\ \citenamefont
		{Milburn}(2010)}]{WisMil10}%
	\BibitemOpen
	\bibfield  {author} {\bibinfo {author} {\bibfnamefont {H.~M.}\ \bibnamefont
			{Wiseman}}\ and\ \bibinfo {author} {\bibfnamefont {G.~J.}\ \bibnamefont
			{Milburn}},\ }\href@noop {} {\textit {\bibinfo {title} {Quantum Measurement and
				Control}}}\ (\bibinfo  {publisher} {Cambridge University Press},\ \bibinfo
	{year} {2010})\BibitemShut {NoStop}%
	\bibitem [{\citenamefont {Braginsky}\ and\ \citenamefont
		{Khalili}(1992)}]{BragKhal}%
	\BibitemOpen
	\bibfield  {author} {\bibinfo {author} {\bibfnamefont {V.}~\bibnamefont
			{Braginsky}}\ and\ \bibinfo {author} {\bibfnamefont {F.}~\bibnamefont
			{Khalili}},\ }\href@noop {} {\textit {\bibinfo {title} {Quantum Measurement}}}\
	(\bibinfo  {publisher} {Cambridge University Press},\ \bibinfo {year}
	{1992})\BibitemShut {NoStop}%
	\bibitem [{\citenamefont {Clerk}\ \textit {et~al.}(2010)\citenamefont {Clerk},
		\citenamefont {Devoret}, \citenamefont {Girvin}, \citenamefont {Marquardt},\
		and\ \citenamefont {Schoelkopf}}]{Clerk10}%
	\BibitemOpen
	\bibfield  {author} {\bibinfo {author} {\bibfnamefont {A.~A.}\ \bibnamefont
			{Clerk}}, \bibinfo {author} {\bibfnamefont {M.~H.}\ \bibnamefont {Devoret}},
		\bibinfo {author} {\bibfnamefont {S.~M.}\ \bibnamefont {Girvin}}, \bibinfo
		{author} {\bibfnamefont {F.}~\bibnamefont {Marquardt}}, \ and\ \bibinfo
		{author} {\bibfnamefont {R.~J.}\ \bibnamefont {Schoelkopf}},\ }\href
	{\doibase 10.1103/RevModPhys.82.1155} {\bibfield  {journal} {\bibinfo
			{journal} {Rev. Mod. Phys.}\ }\textbf {\bibinfo {volume} {82}},\ \bibinfo
		{pages} {1155} (\bibinfo {year} {2010})}\BibitemShut {NoStop}%
	\bibitem [{\citenamefont {{LIGO Scientific Collaboration and Virgo
				Collaboration}}(2016)}]{Abbo016a}%
	\BibitemOpen
	\bibfield  {author} {\bibinfo {author} {\bibnamefont {{LIGO Scientific
					Collaboration and Virgo Collaboration}}},\ }\href {\doibase
		10.1103/PhysRevLett.116.131103} {\bibfield  {journal} {\bibinfo  {journal}
			{Phys. Rev. Lett.}\ }\textbf {\bibinfo {volume} {116}},\ \bibinfo {pages}
		{131103} (\bibinfo {year} {2016})}\BibitemShut {NoStop}%
	\bibitem [{\citenamefont {Brooks}\ \textit {et~al.}(2012)\citenamefont {Brooks},
		\citenamefont {Botter}, \citenamefont {Schreppler}, \citenamefont {Purdy},
		\citenamefont {Brahms},\ and\ \citenamefont {Stamper-Kurn}}]{Stamp12}%
	\BibitemOpen
	\bibfield  {author} {\bibinfo {author} {\bibfnamefont {D.~W.~C.}\
			\bibnamefont {Brooks}}, \bibinfo {author} {\bibfnamefont {T.}~\bibnamefont
			{Botter}}, \bibinfo {author} {\bibfnamefont {S.}~\bibnamefont {Schreppler}},
		\bibinfo {author} {\bibfnamefont {T.~P.}\ \bibnamefont {Purdy}}, \bibinfo
		{author} {\bibfnamefont {N.}~\bibnamefont {Brahms}}, \ and\ \bibinfo {author}
		{\bibfnamefont {D.~M.}\ \bibnamefont {Stamper-Kurn}},\ }\href {\doibase
		10.1038/nature11325} {\bibfield  {journal} {\bibinfo  {journal} {Nature}\
		}\textbf {\bibinfo {volume} {488}},\ \bibinfo {pages} {476} (\bibinfo {year}
		{2012})}\BibitemShut {NoStop}%
	\bibitem [{\citenamefont {Safavi-Naeini}\ \textit
		{et~al.}(2013{\natexlab{a}})\citenamefont {Safavi-Naeini}, \citenamefont
		{Gr{\"o}blacher}, \citenamefont {Hill}, \citenamefont {Chan}, \citenamefont
		{Aspelmeyer},\ and\ \citenamefont {Painter}}]{Pain13}%
	\BibitemOpen
	\bibfield  {author} {\bibinfo {author} {\bibfnamefont {A.~H.}\ \bibnamefont
			{Safavi-Naeini}}, \bibinfo {author} {\bibfnamefont {S.}~\bibnamefont
			{Gr{\"o}blacher}}, \bibinfo {author} {\bibfnamefont {J.~T.}\ \bibnamefont
			{Hill}}, \bibinfo {author} {\bibfnamefont {J.}~\bibnamefont {Chan}}, \bibinfo
		{author} {\bibfnamefont {M.}~\bibnamefont {Aspelmeyer}}, \ and\ \bibinfo
		{author} {\bibfnamefont {O.}~\bibnamefont {Painter}},\ }\href {\doibase
		10.1038/nature12307} {\bibfield  {journal} {\bibinfo  {journal} {Nature}\
		}\textbf {\bibinfo {volume} {500}},\ \bibinfo {pages} {185} (\bibinfo {year}
		{2013}{\natexlab{a}})}\BibitemShut {NoStop}%
	\bibitem [{\citenamefont {Purdy}\ \textit {et~al.}(2013)\citenamefont {Purdy},
		\citenamefont {Yu}, \citenamefont {Peterson}, \citenamefont {Kampel},\ and\
		\citenamefont {Regal}}]{PurReg13}%
	\BibitemOpen
	\bibfield  {author} {\bibinfo {author} {\bibfnamefont {T.~P.}\ \bibnamefont
			{Purdy}}, \bibinfo {author} {\bibfnamefont {P.}~\bibnamefont {Yu}}, \bibinfo
		{author} {\bibfnamefont {R.~W.}\ \bibnamefont {Peterson}}, \bibinfo {author}
		{\bibfnamefont {N.~S.}\ \bibnamefont {Kampel}}, \ and\ \bibinfo {author}
		{\bibfnamefont {C.~A.}\ \bibnamefont {Regal}},\ }\href {\doibase
		10.1103/PhysRevX.3.031012} {\bibfield  {journal} {\bibinfo  {journal} {Phys.
				Rev. X}\ }\textbf {\bibinfo {volume} {3}},\ \bibinfo {pages} {031012}
		(\bibinfo {year} {2013})}\BibitemShut {NoStop}%
	\bibitem [{\citenamefont {Nielsen}\ \textit {et~al.}(2016)\citenamefont
		{Nielsen}, \citenamefont {Tsaturyan}, \citenamefont {Moller}, \citenamefont
		{Polzik},\ and\ \citenamefont {Schliesser}}]{Niel16}%
	\BibitemOpen
	\bibfield  {author} {\bibinfo {author} {\bibfnamefont {W.~H.~P.}\
			\bibnamefont {Nielsen}}, \bibinfo {author} {\bibfnamefont {Y.}~\bibnamefont
			{Tsaturyan}}, \bibinfo {author} {\bibfnamefont {C.~B.}\ \bibnamefont
			{Moller}}, \bibinfo {author} {\bibfnamefont {E.~S.}\ \bibnamefont {Polzik}},
		\ and\ \bibinfo {author} {\bibfnamefont {A.}~\bibnamefont {Schliesser}},\
	}\href {http://arxiv.org/abs/1605.06541} {\bibfield  {journal} {\bibinfo
		{journal} {arXiv:1605.06541}\ } (\bibinfo {year} {2016})}\BibitemShut
{NoStop}%
\bibitem [{\citenamefont {Safavi-Naeini}\ \textit {et~al.}(2012)\citenamefont
	{Safavi-Naeini}, \citenamefont {Chan}, \citenamefont {Hill}, \citenamefont
	{Alegre}, \citenamefont {Krause},\ and\ \citenamefont
	{Painter}}]{AmirPain12}%
\BibitemOpen
\bibfield  {author} {\bibinfo {author} {\bibfnamefont {A.~H.}\ \bibnamefont
		{Safavi-Naeini}}, \bibinfo {author} {\bibfnamefont {J.}~\bibnamefont {Chan}},
	\bibinfo {author} {\bibfnamefont {J.~T.}\ \bibnamefont {Hill}}, \bibinfo
	{author} {\bibfnamefont {T.~P.~M.}\ \bibnamefont {Alegre}}, \bibinfo {author}
	{\bibfnamefont {A.}~\bibnamefont {Krause}}, \ and\ \bibinfo {author}
	{\bibfnamefont {O.}~\bibnamefont {Painter}},\ }\href {\doibase
	10.1103/PhysRevLett.108.033602} {\bibfield  {journal} {\bibinfo  {journal}
		{Phys. Rev. Lett.}\ }\textbf {\bibinfo {volume} {108}},\ \bibinfo {pages}
	{033602} (\bibinfo {year} {2012})}\BibitemShut {NoStop}%
\bibitem [{\citenamefont {Weinstein}\ \textit {et~al.}(2014)\citenamefont
	{Weinstein}, \citenamefont {Lei}, \citenamefont {Wollman}, \citenamefont
	{Suh}, \citenamefont {Metelmann}, \citenamefont {Clerk},\ and\ \citenamefont
	{Schwab}}]{Wein14}%
\BibitemOpen
\bibfield  {author} {\bibinfo {author} {\bibfnamefont {A.}~\bibnamefont
		{Weinstein}}, \bibinfo {author} {\bibfnamefont {C.}~\bibnamefont {Lei}},
	\bibinfo {author} {\bibfnamefont {E.}~\bibnamefont {Wollman}}, \bibinfo
	{author} {\bibfnamefont {J.}~\bibnamefont {Suh}}, \bibinfo {author}
	{\bibfnamefont {A.}~\bibnamefont {Metelmann}}, \bibinfo {author}
	{\bibfnamefont {A.}~\bibnamefont {Clerk}}, \ and\ \bibinfo {author}
	{\bibfnamefont {K.}~\bibnamefont {Schwab}},\ }\href {\doibase
	10.1103/PhysRevX.4.041003} {\bibfield  {journal} {\bibinfo  {journal} {Phys.
			Rev. X}\ }\textbf {\bibinfo {volume} {4}},\ \bibinfo {pages} {041003}
	(\bibinfo {year} {2014})}\BibitemShut {NoStop}%
\bibitem [{\citenamefont {Purdy}\ \textit {et~al.}(2015)\citenamefont {Purdy},
	\citenamefont {Yu}, \citenamefont {Kampel}, \citenamefont {Peterson},
	\citenamefont {Cicak}, \citenamefont {Simmonds},\ and\ \citenamefont
	{Regal}}]{PurReg15}%
\BibitemOpen
\bibfield  {author} {\bibinfo {author} {\bibfnamefont {T.~P.}\ \bibnamefont
		{Purdy}}, \bibinfo {author} {\bibfnamefont {P.-L.}\ \bibnamefont {Yu}},
	\bibinfo {author} {\bibfnamefont {N.~S.}\ \bibnamefont {Kampel}}, \bibinfo
	{author} {\bibfnamefont {R.~W.}\ \bibnamefont {Peterson}}, \bibinfo {author}
	{\bibfnamefont {K.}~\bibnamefont {Cicak}}, \bibinfo {author} {\bibfnamefont
		{R.~W.}\ \bibnamefont {Simmonds}}, \ and\ \bibinfo {author} {\bibfnamefont
		{C.~A.}\ \bibnamefont {Regal}},\ }\href {\doibase 10.1103/PhysRevA.92.031802}
{\bibfield  {journal} {\bibinfo  {journal} {Phys. Rev. A}\ }\textbf {\bibinfo
		{volume} {92}},\ \bibinfo {pages} {031802} (\bibinfo {year}
	{2015})}\BibitemShut {NoStop}%
\bibitem [{\citenamefont {Underwood}\ \textit {et~al.}(2015)\citenamefont
	{Underwood}, \citenamefont {Mason}, \citenamefont {Lee}, \citenamefont {Xu},
	\citenamefont {Jiang}, \citenamefont {Shkarin}, \citenamefont {B{\o}rkje},
	\citenamefont {Girvin},\ and\ \citenamefont {Harris}}]{UndHarr15}%
\BibitemOpen
\bibfield  {author} {\bibinfo {author} {\bibfnamefont {M.}~\bibnamefont
		{Underwood}}, \bibinfo {author} {\bibfnamefont {D.}~\bibnamefont {Mason}},
	\bibinfo {author} {\bibfnamefont {D.}~\bibnamefont {Lee}}, \bibinfo {author}
	{\bibfnamefont {H.}~\bibnamefont {Xu}}, \bibinfo {author} {\bibfnamefont
		{L.}~\bibnamefont {Jiang}}, \bibinfo {author} {\bibfnamefont {A.~B.}\
		\bibnamefont {Shkarin}}, \bibinfo {author} {\bibfnamefont {K.}~\bibnamefont
		{B{\o}rkje}}, \bibinfo {author} {\bibfnamefont {S.~M.}\ \bibnamefont
		{Girvin}}, \ and\ \bibinfo {author} {\bibfnamefont {J.~G.~E.}\ \bibnamefont
		{Harris}},\ }\href {\doibase 10.1103/PhysRevA.92.061801} {\bibfield
	{journal} {\bibinfo  {journal} {Phys. Rev. A}\ }\textbf {\bibinfo {volume}
		{92}},\ \bibinfo {pages} {061801} (\bibinfo {year} {2015})}\BibitemShut
{NoStop}%
\bibitem [{\citenamefont {Shapiro}(1985)}]{Shap85}%
\BibitemOpen
\bibfield  {author} {\bibinfo {author} {\bibfnamefont {J.~H.}\ \bibnamefont
		{Shapiro}},\ }\href
{http://ieeexplore.ieee.org/xpls/abs_all.jsp?arnumber=1072640&tag=1}
{\bibfield  {journal} {\bibinfo  {journal} {IEEE J. Quant. Elec.}\ }\textbf
	{\bibinfo {volume} {21}},\ \bibinfo {pages} {237} (\bibinfo {year}
	{1985})}\BibitemShut {NoStop}%
\bibitem [{\citenamefont {Khalili}\ \textit {et~al.}(2012)\citenamefont
	{Khalili}, \citenamefont {Miao}, \citenamefont {Yang}, \citenamefont
	{Safavi-Naeini}, \citenamefont {Painter},\ and\ \citenamefont
	{Chen}}]{KhalPain12}%
\BibitemOpen
\bibfield  {author} {\bibinfo {author} {\bibfnamefont {F.}~\bibnamefont
		{Khalili}}, \bibinfo {author} {\bibfnamefont {H.}~\bibnamefont {Miao}},
	\bibinfo {author} {\bibfnamefont {H.}~\bibnamefont {Yang}}, \bibinfo {author}
	{\bibfnamefont {A.}~\bibnamefont {Safavi-Naeini}}, \bibinfo {author}
	{\bibfnamefont {O.}~\bibnamefont {Painter}}, \ and\ \bibinfo {author}
	{\bibfnamefont {Y.}~\bibnamefont {Chen}},\ }\href {\doibase
	10.1103/PhysRevA.86.033840} {\bibfield  {journal} {\bibinfo  {journal} {Phys.
			Rev. A}\ }\textbf {\bibinfo {volume} {86}},\ \bibinfo {pages} {033840}
	(\bibinfo {year} {2012})}\BibitemShut {NoStop}%
\bibitem [{\citenamefont {Buchmann}\ \textit {et~al.}(2016)\citenamefont
	{Buchmann}, \citenamefont {Schreppler}, \citenamefont {Kohler}, \citenamefont
	{Spethmann},\ and\ \citenamefont {Stamper-Kurn}}]{Buch16}%
\BibitemOpen
\bibfield  {author} {\bibinfo {author} {\bibfnamefont {L.~F.}\ \bibnamefont
		{Buchmann}}, \bibinfo {author} {\bibfnamefont {S.}~\bibnamefont
		{Schreppler}}, \bibinfo {author} {\bibfnamefont {J.}~\bibnamefont {Kohler}},
	\bibinfo {author} {\bibfnamefont {N.}~\bibnamefont {Spethmann}}, \ and\
	\bibinfo {author} {\bibfnamefont {D.~M.}\ \bibnamefont {Stamper-Kurn}},\
}\href {http://arxiv.org/abs/1602.02141} {\bibfield  {journal} {\bibinfo
	{journal} {arXiv:1602.02141}\ } (\bibinfo {year} {2016})}\BibitemShut
{NoStop}%
\bibitem [{Note1()}]{Note1}%
\BibitemOpen
\bibinfo {note} {Note that sideband asymmetry arising for direct photon
	counting of the meter field has a different origin \cite
	{Wein14}.}\BibitemShut {Stop}%
\bibitem [{\citenamefont {Wilson}\ \textit {et~al.}(2015)\citenamefont {Wilson},
	\citenamefont {Sudhir}, \citenamefont {Piro}, \citenamefont {Schilling},
	\citenamefont {Ghadimi},\ and\ \citenamefont {Kippenberg}}]{WilSudKip15}%
\BibitemOpen
\bibfield  {author} {\bibinfo {author} {\bibfnamefont {D.~J.}\ \bibnamefont
		{Wilson}}, \bibinfo {author} {\bibfnamefont {V.}~\bibnamefont {Sudhir}},
	\bibinfo {author} {\bibfnamefont {N.}~\bibnamefont {Piro}}, \bibinfo {author}
	{\bibfnamefont {R.}~\bibnamefont {Schilling}}, \bibinfo {author}
	{\bibfnamefont {A.}~\bibnamefont {Ghadimi}}, \ and\ \bibinfo {author}
	{\bibfnamefont {T.~J.}\ \bibnamefont {Kippenberg}},\ }\href {\doibase
	10.1038/nature14672} {\bibfield  {journal} {\bibinfo  {journal} {Nature}\
	}\textbf {\bibinfo {volume} {524}},\ \bibinfo {pages} {325} (\bibinfo {year}
	{2015})}\BibitemShut {NoStop}%
\bibitem [{\citenamefont {Hatridge}\ \textit {et~al.}(2013)\citenamefont
	{Hatridge}, \citenamefont {Shankar}, \citenamefont {Mirrahimi}, \citenamefont
	{Shackert}, \citenamefont {Geerlings}, \citenamefont {Brecht}, \citenamefont
	{Sliwa}, \citenamefont {Abdo}, \citenamefont {Frunzio}, \citenamefont
	{Girvin}, \citenamefont {Schoelkopf},\ and\ \citenamefont {Devoret}}]{Hat13}%
\BibitemOpen
\bibfield  {author} {\bibinfo {author} {\bibfnamefont {M.}~\bibnamefont
		{Hatridge}}, \bibinfo {author} {\bibfnamefont {S.}~\bibnamefont {Shankar}},
	\bibinfo {author} {\bibfnamefont {M.}~\bibnamefont {Mirrahimi}}, \bibinfo
	{author} {\bibfnamefont {F.}~\bibnamefont {Shackert}}, \bibinfo {author}
	{\bibfnamefont {K.}~\bibnamefont {Geerlings}}, \bibinfo {author}
	{\bibfnamefont {T.}~\bibnamefont {Brecht}}, \bibinfo {author} {\bibfnamefont
		{K.}~\bibnamefont {Sliwa}}, \bibinfo {author} {\bibfnamefont
		{B.}~\bibnamefont {Abdo}}, \bibinfo {author} {\bibfnamefont {L.}~\bibnamefont
		{Frunzio}}, \bibinfo {author} {\bibfnamefont {S.~M.}\ \bibnamefont {Girvin}},
	\bibinfo {author} {\bibfnamefont {R.~J.}\ \bibnamefont {Schoelkopf}}, \ and\
	\bibinfo {author} {\bibfnamefont {M.~H.}\ \bibnamefont {Devoret}},\ }\href
{\doibase 10.1126/science.1226897} {\bibfield  {journal} {\bibinfo  {journal}
		{Science}\ }\textbf {\bibinfo {volume} {339}},\ \bibinfo {pages} {178}
	(\bibinfo {year} {2013})}\BibitemShut {NoStop}%
\bibitem [{\citenamefont {Taubman}\ \textit {et~al.}(1995)\citenamefont
	{Taubman}, \citenamefont {Wiseman}, \citenamefont {McClelland},\ and\
	\citenamefont {Bachor}}]{Taub95}%
\BibitemOpen
\bibfield  {author} {\bibinfo {author} {\bibfnamefont {M.~S.}\ \bibnamefont
		{Taubman}}, \bibinfo {author} {\bibfnamefont {H.}~\bibnamefont {Wiseman}},
	\bibinfo {author} {\bibfnamefont {D.~E.}\ \bibnamefont {McClelland}}, \ and\
	\bibinfo {author} {\bibfnamefont {H.-A.}\ \bibnamefont {Bachor}},\ }\href
{\doibase 10.1364/JOSAB.12.001792} {\bibfield  {journal} {\bibinfo  {journal}
		{J. Opt. Soc. Am. B}\ }\textbf {\bibinfo {volume} {12}},\ \bibinfo {pages}
	{1792} (\bibinfo {year} {1995})}\BibitemShut {NoStop}%
\bibitem [{\citenamefont {Peterson}\ \textit {et~al.}(2016)\citenamefont
	{Peterson}, \citenamefont {Purdy}, \citenamefont {Kampel}, \citenamefont
	{Andrews}, \citenamefont {Yu},\ and\ \citenamefont {Regal}}]{PetReg15}%
\BibitemOpen
\bibfield  {author} {\bibinfo {author} {\bibfnamefont {R.~W.}\ \bibnamefont
		{Peterson}}, \bibinfo {author} {\bibfnamefont {T.~P.}\ \bibnamefont {Purdy}},
	\bibinfo {author} {\bibfnamefont {N.~S.}\ \bibnamefont {Kampel}}, \bibinfo
	{author} {\bibfnamefont {R.~W.}\ \bibnamefont {Andrews}}, \bibinfo {author}
	{\bibfnamefont {L.~K.~W.}\ \bibnamefont {Yu}, \bibfnamefont {P.-L.}}, \ and\
	\bibinfo {author} {\bibfnamefont {C.~A.}\ \bibnamefont {Regal}},\ }\href
{https://journals.aps.org/prl/abstract/10.1103/PhysRevLett.116.063601}
{\bibfield  {journal} {\bibinfo  {journal} {Phys. Rev. Lett.}\ }\textbf
	{\bibinfo {volume} {116}},\ \bibinfo {pages} {063601} (\bibinfo {year}
	{2016})}\BibitemShut {NoStop}%
\bibitem [{Note2()}]{Note2}%
\BibitemOpen
\bibinfo {note} {Here, heterodyne spectra are expressed as double-sided
	symmetrized spectra, for example $\protect \mathaccentV {bar}016{S}_{yy}$,
	while homodyne spectra are expressed as the corresponding single-sided
	symmetrized versions, for example $\protect \mathaccentV
	{bar}016{S}_y$}\BibitemShut {NoStop}%
\bibitem [{sup()}]{suppinfo}%
\BibitemOpen
\href@noop {} {\bibinfo  {journal} {See Supplementary Information}\
}\BibitemShut {NoStop}%
\bibitem [{\citenamefont {Schilling}\ \textit {et~al.}(2016)\citenamefont
	{Schilling}, \citenamefont {Sch{\"u}tz}, \citenamefont {Ghadimi},
	\citenamefont {Sudhir}, \citenamefont {Wilson},\ and\ \citenamefont
	{Kippenberg}}]{Schil15}%
\BibitemOpen
\bibfield  {journal} {  }\bibfield  {author} {\bibinfo {author} {\bibfnamefont
		{R.}~\bibnamefont {Schilling}}, \bibinfo {author} {\bibfnamefont
		{H.}~\bibnamefont {Sch{\"u}tz}}, \bibinfo {author} {\bibfnamefont
		{A.}~\bibnamefont {Ghadimi}}, \bibinfo {author} {\bibfnamefont
		{V.}~\bibnamefont {Sudhir}}, \bibinfo {author} {\bibfnamefont
		{D.}~\bibnamefont {Wilson}}, \ and\ \bibinfo {author} {\bibfnamefont
		{T.}~\bibnamefont {Kippenberg}},\ }\href {\doibase
	10.1103/PhysRevApplied.5.054019} {\bibfield  {journal} {\bibinfo  {journal}
		{Phys. Rev. Applied}\ }\textbf {\bibinfo {volume} {5}},\ \bibinfo {pages}
	{054019} (\bibinfo {year} {2016})}\BibitemShut {NoStop}%
\bibitem [{\citenamefont {Fabre}\ \textit {et~al.}(1994)\citenamefont {Fabre},
	\citenamefont {Pinard}, \citenamefont {Bourzeix}, \citenamefont {Heidmann},
	\citenamefont {Giacobino},\ and\ \citenamefont {Reynaud}}]{Fabre94}%
\BibitemOpen
\bibfield  {author} {\bibinfo {author} {\bibfnamefont {C.}~\bibnamefont
		{Fabre}}, \bibinfo {author} {\bibfnamefont {M.}~\bibnamefont {Pinard}},
	\bibinfo {author} {\bibfnamefont {S.}~\bibnamefont {Bourzeix}}, \bibinfo
	{author} {\bibfnamefont {A.}~\bibnamefont {Heidmann}}, \bibinfo {author}
	{\bibfnamefont {E.}~\bibnamefont {Giacobino}}, \ and\ \bibinfo {author}
	{\bibfnamefont {S.}~\bibnamefont {Reynaud}},\ }\href
{http://journals.aps.org/pra/abstract/10.1103/PhysRevA.49.1337} {\bibfield
	{journal} {\bibinfo  {journal} {Phys. Rev. A}\ }\textbf {\bibinfo {volume}
		{49}},\ \bibinfo {pages} {1337} (\bibinfo {year} {1994})}\BibitemShut
{NoStop}%
\bibitem [{\citenamefont {Courty}\ \textit {et~al.}(2001)\citenamefont {Courty},
	\citenamefont {Heidmann},\ and\ \citenamefont {Pinard}}]{CouHeid01}%
\BibitemOpen
\bibfield  {author} {\bibinfo {author} {\bibfnamefont {J.}~\bibnamefont
		{Courty}}, \bibinfo {author} {\bibfnamefont {A.}~\bibnamefont {Heidmann}}, \
	and\ \bibinfo {author} {\bibfnamefont {M.}~\bibnamefont {Pinard}},\ }\href
{http://www.springerlink.com/index/5LQFX1PKY0AAEK6A.pdf} {\bibfield
	{journal} {\bibinfo  {journal} {Eur. Phys. J. D}\ }\textbf {\bibinfo {volume}
		{17}},\ \bibinfo {pages} {399} (\bibinfo {year} {2001})}\BibitemShut
{NoStop}%
\bibitem [{\citenamefont {Jayich}\ \textit {et~al.}(2012)\citenamefont {Jayich},
	\citenamefont {Sankey}, \citenamefont {B{\o}rkje}, \citenamefont {Lee},
	\citenamefont {Yang}, \citenamefont {Underwood}, \citenamefont {Childress},
	\citenamefont {Petrenko}, \citenamefont {Girvin},\ and\ \citenamefont
	{Harris}}]{JayHarris12}%
\BibitemOpen
\bibfield  {author} {\bibinfo {author} {\bibfnamefont {A.~M.}\ \bibnamefont
		{Jayich}}, \bibinfo {author} {\bibfnamefont {J.~C.}\ \bibnamefont {Sankey}},
	\bibinfo {author} {\bibfnamefont {K.}~\bibnamefont {B{\o}rkje}}, \bibinfo
	{author} {\bibfnamefont {D.}~\bibnamefont {Lee}}, \bibinfo {author}
	{\bibfnamefont {C.}~\bibnamefont {Yang}}, \bibinfo {author} {\bibfnamefont
		{M.}~\bibnamefont {Underwood}}, \bibinfo {author} {\bibfnamefont
		{L.}~\bibnamefont {Childress}}, \bibinfo {author} {\bibfnamefont
		{A.}~\bibnamefont {Petrenko}}, \bibinfo {author} {\bibfnamefont {S.~M.}\
		\bibnamefont {Girvin}}, \ and\ \bibinfo {author} {\bibfnamefont {J.~G.~E.}\
		\bibnamefont {Harris}},\ }\href {\doibase 10.1088/1367-2630/14/11/115018}
{\bibfield  {journal} {\bibinfo  {journal} {New J. Phys.}\ }\textbf {\bibinfo
		{volume} {14}},\ \bibinfo {pages} {115018} (\bibinfo {year}
	{2012})}\BibitemShut {NoStop}%
\bibitem [{\citenamefont {Wiseman}(1999)}]{Wise99}%
\BibitemOpen
\bibfield  {author} {\bibinfo {author} {\bibfnamefont {H.~M.}\ \bibnamefont
		{Wiseman}},\ }\href {\doibase 10.1088/1464-4266/1/4/317} {\bibfield
	{journal} {\bibinfo  {journal} {J. Opt. B}\ }\textbf {\bibinfo {volume}
		{1}},\ \bibinfo {pages} {459} (\bibinfo {year} {1999})}\BibitemShut {NoStop}%
\bibitem [{\citenamefont {Safavi-Naeini}\ \textit
	{et~al.}(2013{\natexlab{b}})\citenamefont {Safavi-Naeini}, \citenamefont
	{Chan}, \citenamefont {Hill}, \citenamefont {Gr{\"o}blacher}, \citenamefont
	{Miao}, \citenamefont {Chen}, \citenamefont {Aspelmeyer},\ and\ \citenamefont
	{Painter}}]{AmirPain13}%
\BibitemOpen
\bibfield  {author} {\bibinfo {author} {\bibfnamefont {A.~H.}\ \bibnamefont
		{Safavi-Naeini}}, \bibinfo {author} {\bibfnamefont {J.}~\bibnamefont {Chan}},
	\bibinfo {author} {\bibfnamefont {J.~T.}\ \bibnamefont {Hill}}, \bibinfo
	{author} {\bibfnamefont {S.}~\bibnamefont {Gr{\"o}blacher}}, \bibinfo
	{author} {\bibfnamefont {H.}~\bibnamefont {Miao}}, \bibinfo {author}
	{\bibfnamefont {Y.}~\bibnamefont {Chen}}, \bibinfo {author} {\bibfnamefont
		{M.}~\bibnamefont {Aspelmeyer}}, \ and\ \bibinfo {author} {\bibfnamefont
		{O.}~\bibnamefont {Painter}},\ }\href {\doibase
	10.1088/1367-2630/15/3/035007} {\bibfield  {journal} {\bibinfo  {journal}
		{New J. Phys.}\ }\textbf {\bibinfo {volume} {15}},\ \bibinfo {pages} {035007}
	(\bibinfo {year} {2013}{\natexlab{b}})}\BibitemShut {NoStop}%
\bibitem [{Note3()}]{Note3}%
\BibitemOpen
\bibinfo {note} {In addition, it is known that for semiconductor lasers,
	phase-amplitude correlations are limited to frequencies close to their
	relaxation oscillation frequency \cite {Vah83,Exet92}; the latter is
	typically at a few GHz from the carrier \cite {Kip13} -- irrelevant for our
	experiment}\BibitemShut {NoStop}%
\bibitem [{\citenamefont {Aspelmeyer}\ \textit {et~al.}(2014)\citenamefont
	{Aspelmeyer}, \citenamefont {Kippenberg},\ and\ \citenamefont
	{Marquardt}}]{AspKipMar14}%
\BibitemOpen
\bibfield  {author} {\bibinfo {author} {\bibfnamefont {M.}~\bibnamefont
		{Aspelmeyer}}, \bibinfo {author} {\bibfnamefont {T.~J.}\ \bibnamefont
		{Kippenberg}}, \ and\ \bibinfo {author} {\bibfnamefont {F.}~\bibnamefont
		{Marquardt}},\ }\href {\doibase 10.1103/RevModPhys.86.1391} {\bibfield
	{journal} {\bibinfo  {journal} {Rev. Mod. Phys.}\ }\textbf {\bibinfo {volume}
		{86}},\ \bibinfo {pages} {1391} (\bibinfo {year} {2014})}\BibitemShut
{NoStop}%
\bibitem [{\citenamefont {Gardiner}\ and\ \citenamefont
	{Collett}(1985)}]{Gard85}%
\BibitemOpen
\bibfield  {author} {\bibinfo {author} {\bibfnamefont {C.}~\bibnamefont
		{Gardiner}}\ and\ \bibinfo {author} {\bibfnamefont {M.}~\bibnamefont
		{Collett}},\ }\href {\doibase 10.1103/PhysRevA.31.3761} {\bibfield  {journal}
	{\bibinfo  {journal} {Phys. Rev. A}\ }\textbf {\bibinfo {volume} {31}},\
	\bibinfo {pages} {3761} (\bibinfo {year} {1985})}\BibitemShut {NoStop}%
\bibitem [{\citenamefont {Gorodetsky}\ and\ \citenamefont
	{Grudinin}(2004)}]{Gor04}%
\BibitemOpen
\bibfield  {author} {\bibinfo {author} {\bibfnamefont {M.~L.}\ \bibnamefont
		{Gorodetsky}}\ and\ \bibinfo {author} {\bibfnamefont {I.~S.}\ \bibnamefont
		{Grudinin}},\ }\href {\doibase 10.1364/JOSAB.21.000697} {\bibfield  {journal}
	{\bibinfo  {journal} {JOSA B}\ }\textbf {\bibinfo {volume} {21}},\ \bibinfo
	{pages} {697} (\bibinfo {year} {2004})}\BibitemShut {NoStop}%
\bibitem [{\citenamefont {Gillespie}\ and\ \citenamefont
	{Raab}(1993)}]{Gill93}%
\BibitemOpen
\bibfield  {author} {\bibinfo {author} {\bibfnamefont {A.}~\bibnamefont
		{Gillespie}}\ and\ \bibinfo {author} {\bibfnamefont {F.}~\bibnamefont
		{Raab}},\ }\href
{http://www.sciencedirect.com/science/article/pii/037596019390861S}
{\bibfield  {journal} {\bibinfo  {journal} {Phys. Lett. A}\ }\textbf
	{\bibinfo {volume} {178}},\ \bibinfo {pages} {357} (\bibinfo {year}
	{1993})}\BibitemShut {NoStop}%
\bibitem [{\citenamefont {Bjorklund}(1980)}]{Bjork80}%
\BibitemOpen
\bibfield  {author} {\bibinfo {author} {\bibfnamefont {G.~C.}\ \bibnamefont
		{Bjorklund}},\ }\href {\doibase 10.1364/OL.5.000015} {\bibfield  {journal}
	{\bibinfo  {journal} {Opt. Lett.}\ }\textbf {\bibinfo {volume} {5}},\
	\bibinfo {pages} {15} (\bibinfo {year} {1980})}\BibitemShut {NoStop}%
\bibitem [{\citenamefont {Gorodetsky}\ \textit {et~al.}(2010)\citenamefont
	{Gorodetsky}, \citenamefont {Schliesser}, \citenamefont {Anetsberger},
	\citenamefont {Deleglise},\ and\ \citenamefont {Kippenberg}}]{GorKip10}%
\BibitemOpen
\bibfield  {author} {\bibinfo {author} {\bibfnamefont {M.~L.}\ \bibnamefont
		{Gorodetsky}}, \bibinfo {author} {\bibfnamefont {A.}~\bibnamefont
		{Schliesser}}, \bibinfo {author} {\bibfnamefont {G.}~\bibnamefont
		{Anetsberger}}, \bibinfo {author} {\bibfnamefont {S.}~\bibnamefont
		{Deleglise}}, \ and\ \bibinfo {author} {\bibfnamefont {T.~J.}\ \bibnamefont
		{Kippenberg}},\ }\href {\doibase 10.1126/science.1195596} {\bibfield
	{journal} {\bibinfo  {journal} {Opt. Exp.}\ }\textbf {\bibinfo {volume}
		{18}},\ \bibinfo {pages} {23236} (\bibinfo {year} {2010})}\BibitemShut
{NoStop}%
\bibitem [{\citenamefont {Braginsky}\ \textit {et~al.}(2000)\citenamefont
	{Braginsky}, \citenamefont {Gorodetsky},\ and\ \citenamefont
	{Vyatchanin}}]{BragGor00}%
\BibitemOpen
\bibfield  {author} {\bibinfo {author} {\bibfnamefont {V.~B.}\ \bibnamefont
		{Braginsky}}, \bibinfo {author} {\bibfnamefont {M.~L.}\ \bibnamefont
		{Gorodetsky}}, \ and\ \bibinfo {author} {\bibfnamefont {S.~P.}\ \bibnamefont
		{Vyatchanin}},\ }\href {\doibase 10.1016/S0375-9601(00)00389-3} {\bibfield
	{journal} {\bibinfo  {journal} {Phys. Lett. A}\ }\textbf {\bibinfo {volume}
		{271}},\ \bibinfo {pages} {303} (\bibinfo {year} {2000})}\BibitemShut
{NoStop}%
\bibitem [{\citenamefont {Cramer}()}]{Cram}%
\BibitemOpen
\bibfield  {author} {\bibinfo {author} {\bibfnamefont {H.}~\bibnamefont
		{Cramer}},\ }\href@noop {} {\textit {\bibinfo {title} {Mathematical Methods of
			Statistics}}}\ (\bibinfo  {publisher} {Princeton University
	Press})\BibitemShut {NoStop}%
\bibitem [{\citenamefont {Simon}\ \textit {et~al.}(1994)\citenamefont {Simon},
	\citenamefont {Mukunda},\ and\ \citenamefont {Dutta}}]{SimMuk94}%
\BibitemOpen
\bibfield  {author} {\bibinfo {author} {\bibfnamefont {R.}~\bibnamefont
		{Simon}}, \bibinfo {author} {\bibfnamefont {N.}~\bibnamefont {Mukunda}}, \
	and\ \bibinfo {author} {\bibfnamefont {B.}~\bibnamefont {Dutta}},\ }\href
{http://link.aps.org/doi/10.1103/PhysRevA.49.1567} {\bibfield  {journal}
	{\bibinfo  {journal} {Phys. Rev. A}\ }\textbf {\bibinfo {volume} {49}},\
	\bibinfo {pages} {1567} (\bibinfo {year} {1994})}\BibitemShut {NoStop}%
\bibitem [{\citenamefont {Vahala}\ \textit {et~al.}(1983)\citenamefont {Vahala},
	\citenamefont {Harder},\ and\ \citenamefont {Yariv}}]{Vah83}%
\BibitemOpen
\bibfield  {author} {\bibinfo {author} {\bibfnamefont {K.}~\bibnamefont
		{Vahala}}, \bibinfo {author} {\bibfnamefont {C.}~\bibnamefont {Harder}}, \
	and\ \bibinfo {author} {\bibfnamefont {A.}~\bibnamefont {Yariv}},\ }\href
{http://scitation.aip.org/content/aip/journal/apl/42/3/10.1063/1.93894}
{\bibfield  {journal} {\bibinfo  {journal} {Appl. Phys. Lett.}\ }\textbf
	{\bibinfo {volume} {42}},\ \bibinfo {pages} {211} (\bibinfo {year}
	{1983})}\BibitemShut {NoStop}%
\bibitem [{\citenamefont {van Exeter}\ \textit {et~al.}(1992)\citenamefont {van
		Exeter}, \citenamefont {Hamel}, \citenamefont {Woerdman},\ and\ \citenamefont
	{Zeijlmans}}]{Exet92}%
\BibitemOpen
\bibfield  {author} {\bibinfo {author} {\bibfnamefont {M.}~\bibnamefont {van
			Exeter}}, \bibinfo {author} {\bibfnamefont {W.}~\bibnamefont {Hamel}},
	\bibinfo {author} {\bibfnamefont {J.~P.}\ \bibnamefont {Woerdman}}, \ and\
	\bibinfo {author} {\bibfnamefont {B.}~\bibnamefont {Zeijlmans}},\ }\href
{http://ieeexplore.ieee.org/xpls/abs_all.jsp?arnumber=135299} {\bibfield
	{journal} {\bibinfo  {journal} {IEEE J. Quantum Electronics}\ }\textbf
	{\bibinfo {volume} {28}},\ \bibinfo {pages} {1470} (\bibinfo {year}
	{1992})}\BibitemShut {NoStop}%
\bibitem [{\citenamefont {Kippenberg}\ \textit {et~al.}(2013)\citenamefont
	{Kippenberg}, \citenamefont {Schliesser},\ and\ \citenamefont
	{Gorodetsky}}]{Kip13}%
\BibitemOpen
\bibfield  {author} {\bibinfo {author} {\bibfnamefont {T.~J.}\ \bibnamefont
		{Kippenberg}}, \bibinfo {author} {\bibfnamefont {A.}~\bibnamefont
		{Schliesser}}, \ and\ \bibinfo {author} {\bibfnamefont {M.~L.}\ \bibnamefont
		{Gorodetsky}},\ }\href
{http://iopscience.iop.org/article/10.1088/1367-2630/15/1/015019} {\bibfield
	{journal} {\bibinfo  {journal} {New J. Phys.}\ }\textbf {\bibinfo {volume}
		{15}},\ \bibinfo {pages} {015019} (\bibinfo {year} {2013})}\BibitemShut
{NoStop}%
\end{thebibliography}
\end{document}